\documentclass[aps,prd,eqsecnum,showpacs,floatfix,twocolumn]{revtex4}
\usepackage{latexsym}
\usepackage{graphicx,subfigure,amsthm}% Include figure files
\newtheorem*{The*}{Theorem}

\begin{document}
%\draft

\thispagestyle{empty}

\title{
Non-existence of self-similar solutions containing a black hole in
a universe with a stiff fluid or scalar field or quintessence}
\author{Tomohiro Harada$^{1,2,}$\footnote{Electronic address:harada@rikkyo.ac.jp},
Hideki Maeda$^{1,3,4,}$\footnote{Electronic address:hideki@gravity.phys.waseda.ac.jp}
and
B.~J.~Carr$^{5,6,}$\footnote{Electronic address:B.J.Carr@qmul.ac.uk}}
\affiliation{
$^{1}$Department of Physics, Rikkyo University, Tokyo 171-8501, Japan, 
$^{2}$Department of Physics, Kyoto University, Kyoto 606-8502, Japan, 
$^{3}$Graduate School of Science and Engineering, Waseda University,
Tokyo 169-8555, Japan, 
$^{4}$Department of Physics, International Christian University, 
Mitaka, Tokyo 181-8585, Japan,
$^{5}$Research Center for the Early Universe, Graduate School of Science, 
University of Tokyo, Tokyo 113-0033, Japan, 
$^{6}$Astronomy Unit, Queen Mary, University of London, Mile End Road, London E1 4NS, UK}
\date{\today}

\begin{abstract}                % DON'T CHANGE THIS LINE
We consider the possible existence of self-similar solutions containing
 black holes in a Friedmann background with a stiff fluid or a scalar
 field. We carefully study the relationship between the self-similar
 equations in these two cases and emphasize the crucial role of the
 similarity horizon. We show that there is no self-similar  black hole
 solution surrounded by an exact or asymptotically flat Friedmann
 background containing a massless scalar field. 
This result also applies for a scalar field with a potential, providing
 the universe is decelerating.  However, if there is a potential and the
 universe is accelerating (as in the quintessence scenario), the result
 only applies for an exact Friedmann background. 
This extends the result previously found in the stiff fluid case and strongly suggests that accretion onto 
primordial black holes is ineffective even during scalar field domination. 
It also contradicts recent claims that such black holes can grow appreciably by accreting quintessence. Appreciable growth might be possible with very special matter fields but this requires {\it ad hoc} and probably unphysical conditions.
\end{abstract}
\maketitle
%\tableofcontents

\section{Introduction}
Scalar fields are one of the key ingredients  in modern cosmology.
They play an important role during 
inflation and certain phase transitions; they feature in the preheating and 
quintessence scenarios; and they are pervasive in all sorts of alternative theories (eg. string theory and scalar-tensor theory) which are likely to be relevant in the high curvature phase of the early Universe.
In such scenarios, 
it is often assumed that scalar fields dominated 
the energy of the Universe at some stage of its evolution.  This could have profound implications for the formation and evolution of any primordial black holes (PBHs) that may have formed in the early 
Universe~\cite{hawking1971}. It would also modify the constraints on the number of 
PBHs provided by various observations~\cite{carr1975,carr2003}.

A particularly interesting example of this arises if the dark energy which generates the cosmic acceleration observed at the present epoch is associated with a quintessence field. In this context, it has recently been claimed~\cite{bm2002,ch2005} 
that PBHs could grow enough through accreting quintessence to provide seeds for the sort of supermassive black holes thought to reside in galactic nuclei~\cite{korm1995}.
This claim is based on a simple Newtonian analysis, in which a PBH with size comparable to the cosmological particle horizon appears to grow in a self-similar manner~\cite{zn1967}, i.e. its size is always the same fraction of the size of the particle horizon.  This suggests that PBHs may accrete surrounding mass very
effectively if their initial size is fine-tuned. However, this analysis neglects the cosmological expansion
~\cite{ch1974,hc2005a}. 
More recently,  a general relativistic analysis of this problem has been applied for a black hole accreting 
dark energy and phantom energy~\cite{bde2004,bde2005,no2004},
as well as a ghost condensate~\cite{frolov2004,mukohyama2005}. 
However, the cosmological expansion is still neglected in these analyses.

The possibility of spherically symmetric self-similar PBH 
solutions was first studied more than 30 years ago for a fluid 
with a radiation equation of state ($p=\epsilon/3$)~\cite{ch1974}.  This study was also prompted by the Newtonian analysis and demonstrated analytically  that 
there is no self-similar solution which contains a black 
hole attached to an exact Friedmann background via a sonic point (i.e. in which the black hole forms by purely causal processes). The Newtonian analysis is therefore definitely misleading in this case. There are self-similar solutions which are {\it asymptotically} Friedmann at large distances from the black hole. However, these correspond to special initial conditions, in which matter is effectively thrown into the black hole at every distance; they do not contain a sonic point because they are supersonic everywhere. 

Similar results were subsequently proved for all perfect fluid systems with equation of state of the form $p=k\epsilon$ with $0 < k < 1$~\cite{carr1976,bh1978b,carryahil1990}. Indeed it was shown that the only physical self-similar solution which can be attached to an external Friedmann solution via a sonic point is Friedmann itself; the other solutions either enter a negative mass regime or encounter another sonic point where the pressure gradient diverges~\cite{bh1978b}. Later numerical relativistic calculations have confirmed that PBHs cannot 
accrete much for these perfect fluid 
systems~\cite{nnp1978,np1980,nj1999,jn1999,ss1999,hs2002,mmr2005}. It has also been shown that the asymptotically Friedmann self-similar solutions with black holes are not strictly Friedmann at infinity, because they exhibit a solid angle deficit which could in principle show up in the angular diameter test~\cite{mkm2002}.

The limiting values of $k$ require special
consideration. In the $k=0$ case, there is no sonic point, so the uniqueness is trivial; again there are asymptotically Friedmann self-similar solutions with PBHs but only for special initial conditions~\cite{ch1974}. The stiff fluid case ($k=1$) is more complicated and has an interesting history. It was originally claimed by Lin et al.~\cite{lcf1976} that there {\it could} exist a
self-similar black hole solution in this case but 
subsequent calculations by Bicknell and Henriksen~\cite{bh1978a} showed that 
this solution does not contain a black hole after all
because the alleged event horizon is timelike. The only way to avoid this conclusion would be
if the fluid turned into null dust at this timelike surface,
but this seems physically implausible. It therefore appears that there is no self-similar solution containing a black hole in an exact Friedmann background for any value of $k$ in the range [0,1]. 

In view of the prevalence of scalar fields in the early Universe, it is clearly important  to know whether the non-existence of a self-similar solution extends to the scalar field case. One might expect this from the stiff fluid analysis, since a scalar field is equivalent to a stiff fluid provided the gradient is everywhere timelike~\cite{mad1988}. This is supported by recent numerical studies of the growth of PBHs in a scalar field system, which -  for a variety of initial conditions - give no evidence for self-similar growth~\cite{hc2005b,hc2005c}.
However, it is very difficult to determine through numerical simulations 
whether self-similar growth is impossible 
for any specific initial data. One therefore needs an  
{\it analytic} proof that there is no self-similar 
black hole solution if the cosmological expansion is taken
into account. This paper presents such a proof and also provides the basis for studying
the growth of black holes in an expanding universe 
containing more general matter fields. 

The plan of this paper is as follows. In Section II, we review the stiff fluid case, with particular emphasis on the earlier work of Lin et al.~\cite{lcf1976} and Bicknell and Henriksen~\cite{bh1978a}. In Section III, we consider the massless scalar field case, using Brady's 
formulation of the self-similar equations
as an autonomous system~\cite{brady1995} and making a detailed comparison with the stiff fluid case. We prove that there is no self-similar solution unless the behaviour of the matter is very contrived. In Section IV we extend the analysis to the case of a scalar field with an exponential potential (as applies in the quintessence scenario) and show that the proof applies in this case also. We summarize our conclusions in Section V. Various technical issues are relegated to appendices, including a comparison with previous literature, a discussion of the global characteristics of these solutions, and an analysis of the behaviour of solutions at the similarity horizon. We use geometrised units with $c=G=1$ throughout this paper.

\section{Stiff fluid case}
\label{sec:stiff_fluid}
\subsection{Bicknell and Henriksen's formulation}
First, we show how a stiff fluid
description of a scalar field can give insight into the 
fluid-field correspondence for self-similar 
solutions~\cite{hm2003}.
A stiff fluid is a perfect fluid with the equation of state
$p=\epsilon$ and stress-energy tensor 
\begin{equation}
T^{ab}=\epsilon (2u^{a}u^{b}+g^{ab}).
\end{equation}
If $u^{a}$ is vorticity-free, 
one can show that this matter field is equivalent to a massless
scalar field $\phi$, 
for which the stress-energy tensor is
\begin{equation}
T_{ab}=\phi_{,a}\phi_{,b}-\frac12 g_{ab} \phi_{,c}\phi^{,c}
\label{scalarstress}
\end{equation}
and the gradient $\phi_{,a}$ is timelike. The associated energy density and 4-velocity are
\begin{equation}
\epsilon = -\frac{1}{2}\phi^{,c}\phi_{,c},\quad 
u_{a}=\pm \frac{\phi_{,a}}{\sqrt{-\phi_{,c}\phi^{,c}}},
\label{eq:epsilon}
\end{equation}
where the sign in the second equation is chosen so that $u^{a}$ is future-directed.
The line element in a spherically symmetric spacetime can be written as
\begin{equation}
ds^{2}=-e^{2\nu(t,R)}dt^{2}+e^{2\lambda(t,R)}dR^{2}+r^{2}(t,R)(d\theta^{2}+\sin^{2}\theta
 d\phi^{2}),
\end{equation}
where $R$ and $r$ are  
the comoving and area radial coordinates, respectively.

Let us now assume self-similarity (i.e. the homothetic condition),
which implies that we can write 
\begin{equation}
\nu=\nu(X),\lambda=\lambda(X), S(X)=\frac{r}{R}, \phi=h(X)-\kappa \ln |t|,
\label{eq:phi_kappa}
\end{equation}
where $X \equiv \ln (R/|t|)$. 
The constant $\kappa$ is arbitrary 
but we will need to assume $\kappa =1/\sqrt{12\pi}$ below in order to allow the Friedmann solution.
If we adopt comoving coordinates, which is always possible providing the 4-velocity is timelike, then $T^{\mu}_{\nu}$
is diagonal. In this case, $h(X)$ must be constant and so $\phi$ depends only on $t$. 

Following Ref.~\cite{bh1978b}, we now introduce the following variables:
\begin{equation}
V^{2}=e^{2\lambda-2\nu+2X}, \quad E =8\pi \epsilon R^{2},\quad
A=\frac{m}{4\pi \epsilon r^{3}}\, ,
\label{eq:V2EA}
\end{equation}
where $V$ is the speed of the fluid flow relative to 
the similarity surface $X$=const, $m$ is the Misner-Sharp mass, $E$ is a
dimensionless measure of the energy density and $A$ is one third of the
ratio of the average density within $r$ to the local density there. 
Energy conservation yields the following
integrals:
\begin{equation}
e^{2\nu}=a_{\nu}E^{-1}e^{2X}, \quad
e^{2\lambda}=a_{\lambda}E^{-1}S^{-4}, 
\label{eq:enuelambda}
\end{equation}
where $a_{\nu}$ and $a_{\lambda}$ are constants of integration.
Eqs.~(\ref{eq:epsilon}), (\ref{eq:phi_kappa}) and (\ref{eq:V2EA})
give
\begin{equation}
e^{2\nu}=4\pi \kappa^{2} E^{-1} e^{2X},
\label{eq:e2nu}
\end{equation}
so $a_{\nu}$ is identified as $4\pi \kappa^{2}$
and is invariant under rescaling of $t$ and $R$. 
If we rescale $R$ appropriately, we can also
set $a_{\lambda}$ to be $4\pi\kappa^{2}$. 
Eqs.~(\ref{eq:V2EA}) and (\ref{eq:enuelambda})
then imply~\cite{bh1978a} 
\begin{equation}
V^{2}=S^{-4},
\end{equation}
so the Einstein field equations
reduce to the following autonomous system~\cite{bh1978a,hm2003}:
\begin{eqnarray}
(V^{2})'&=&2(1-A)V^{2}, 
\label{eq:V2'}\\ 
A'&=&\frac{1}{2}(1+A)(1-3 A) \nonumber \\
&&-2 A V^{2}\frac{A-4\pi \kappa^{2}}{1-V^{2}}, 
\label{eq:A'}\\
E'&=&2 \left[V^{2}\frac{A-4\pi\kappa^{2}}{1-V^{2}}+1 \right]E,
\label{eq:eta'}
\end{eqnarray}
together with a constraint equation
\begin{equation}
V^{2}(1-A)^{2}-(1+A)^{2}+16\pi\kappa^{2}V^{2}(E^{-1}|V|-A)=0.
\label{eq:const1}
\end{equation}
Here a prime denotes a derivative with respect to $X$.

\subsection{$V^{2}=1$ surface}
Once we have determined  initial values satisfying
Eq.~(\ref{eq:const1}), Eqs.~(\ref{eq:V2'}) and (\ref{eq:A'}) constitute
a two-dimensional autonomous system and Eq.~(\ref{eq:eta'}) can be
evolved freely.
However, Eq.~(\ref{eq:A'}) shows that the condition $V^{2}=1$ corresponds to a singular point in the above system
of differential equations,
so the usual uniqueness of solutions may not apply.
This is because one may have a sonic point where $V^{2}=1$ and, in the
stiff fluid case, this coincides with a null surface;
we term this a ``similarity horizon''.

\begin{figure}
\includegraphics[scale=1]{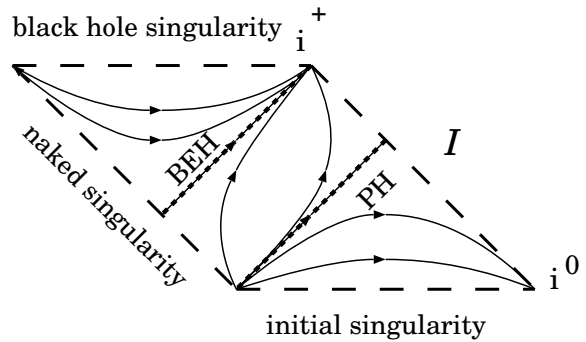}
\caption{\label{fig:dec_pbh_penrose}
The causal structure of self-similar black holes 
in a decelerating universe.
The trajectories of similarity surfaces are shown.
There is a spacelike big bang singularity, a null naked singularity and
a spacelike black hole singularity.
The particle horizon (PH) and black hole event horizon (BEH) are both
similarity horizons.}
\end{figure}

In order to understand the significance of the similarity horizon, it is
useful to draw the Penrose diagram for the self-similar black hole 
spacetime. This
is shown in Fig.~\ref{fig:dec_pbh_penrose}, which
is also equivalent to Figure 5 of~\cite{cg2003}. It can be obtained by
combining the Penrose diagrams for a decelerating flat Friedmann universe,
which is discussed in Appendices~\ref{sec:friedmann}
and \ref{sec:friedmann_self-similar}, and that for 
a naked singularity~\cite{op1990}.
The spacetime is asymptotically flat Friedmann with a spacelike big bang
singularity outside the particle horizon. It also has a null naked
singularity, a spacelike black hole singularity and a black hole event horizon.
The similarity surfaces are spacelike outside the particle horizon
and inside the black hole event horizon, timelike in between 
and null on the horizons themselves. This shows that the 
particle horizon and black hole event horizon both correspond to 
similarity horizons.

We note that the induced metric on the $X=\mbox{const}$ surface
can be written as
\begin{equation}
ds^{2}=-e^{2\nu}(1-V^{2})dt^{2}+r^{2}(d\theta^{2}+\sin^{2}\theta
 d\phi^{2}).
\label{eq:xi_const_surface}
\end{equation}
If this surface is null, it can be identified with
the black hole event horizon or the cosmological particle horizon, both corresponding to the condition
$V^{2}=1$ providing $\nu$ remains finite.
If the gradient of the energy density is finite at the similarity
horizon, Eq.~(\ref{eq:eta'}) also requires $A=4\pi \kappa^{2}$.
This coincidence between singular points of the ordinary
differential equations (ODEs), i.e. sonic points, and
similarity horizons only applies for a stiff fluid.
Sonic points are timelike surfaces for perfect fluids
with the equation of state $p=k \epsilon$ ($0< k <1$).

\begin{figure}[htbp]
\includegraphics[scale=0.8]{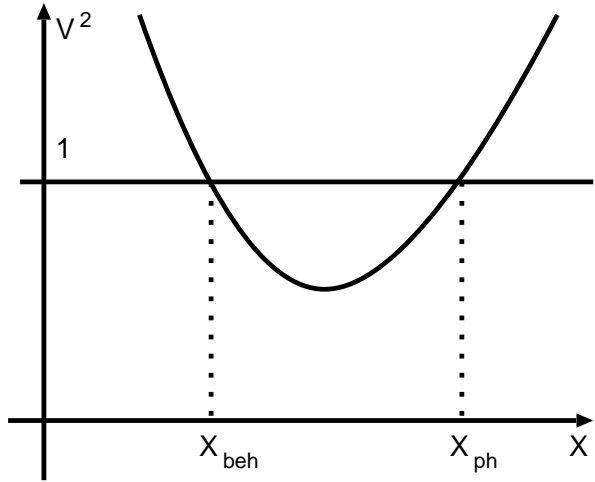}
\caption{\label{fig:expected_V2}
The required behaviour of the velocity function $V^{2}(X)$
for a black hole solution which is attached to an exact Friedmann background.}
\end{figure}

If one wants a solution which represents a black hole surrounded
by the exact Friedmann solution, we first note that the velocity
gradient must be positive on both sides of the matching surface. 
(This is a consequence of the fact that the Friedmann sonic point is a node, as discussed in 
Section~\ref{subsec:structure_similarity_horizon}.) 
Therefore $V^{2}(X)$ must have the behaviour shown 
in Fig.~\ref{fig:expected_V2}. 
As $X$ decreases from infinity, 
$V^{2}$ first decreases, goes below 1, reaches a minimum, then 
increases and reaches 1 again.
The first point where $V^2$ crosses 1 is the 
particle horizon, $X=X_{\rm ph}$. One might expect the second point where
$V^2$ crosses 1 to be the black hole event horizon, $X=X_{\rm beh}$, and indeed Lin et al.~\cite{lcf1976} make this interpretation.
However, Bicknell and Henriksen~\cite{bh1978a} showed that this is not the case: the similarity surface is timelike rather than null because $\nu$ diverges there. More precisely, 
$E$ goes to $0$, which shows that the gradient of 
the scalar field is null, and the factor $e^{2\nu}(1-V^{2})$ in Eq.~(\ref{eq:xi_const_surface}) tends to a finite positive limit,
which shows that the second crossing surface 
is timelike.
This reveals the limitation of the comoving-coordinate 
framework for scalar field systems. The Lin et al. and Bicknell and Henriksen analyses are discussed in more detail in Appendix~\ref{sec:otherwork}. 

Because Bicknell and Henriksen could only find a solution in which 
$V^{2}$ has the required behaviour up to the point where $V^{2}=1$, they inferred that no black hole solution is possible if the fluid has the same equation of state everywhere. However, they did find that  a black hole solution is possible if the stiff fluid turns into null dust at the surface where $V^{2}=1$. This situation is obviously contrived, so in
this paper we seek a more physical extension with a massless scalar field. 

\section{Massless scalar field case}

\begin{figure*}
\includegraphics[scale=1]{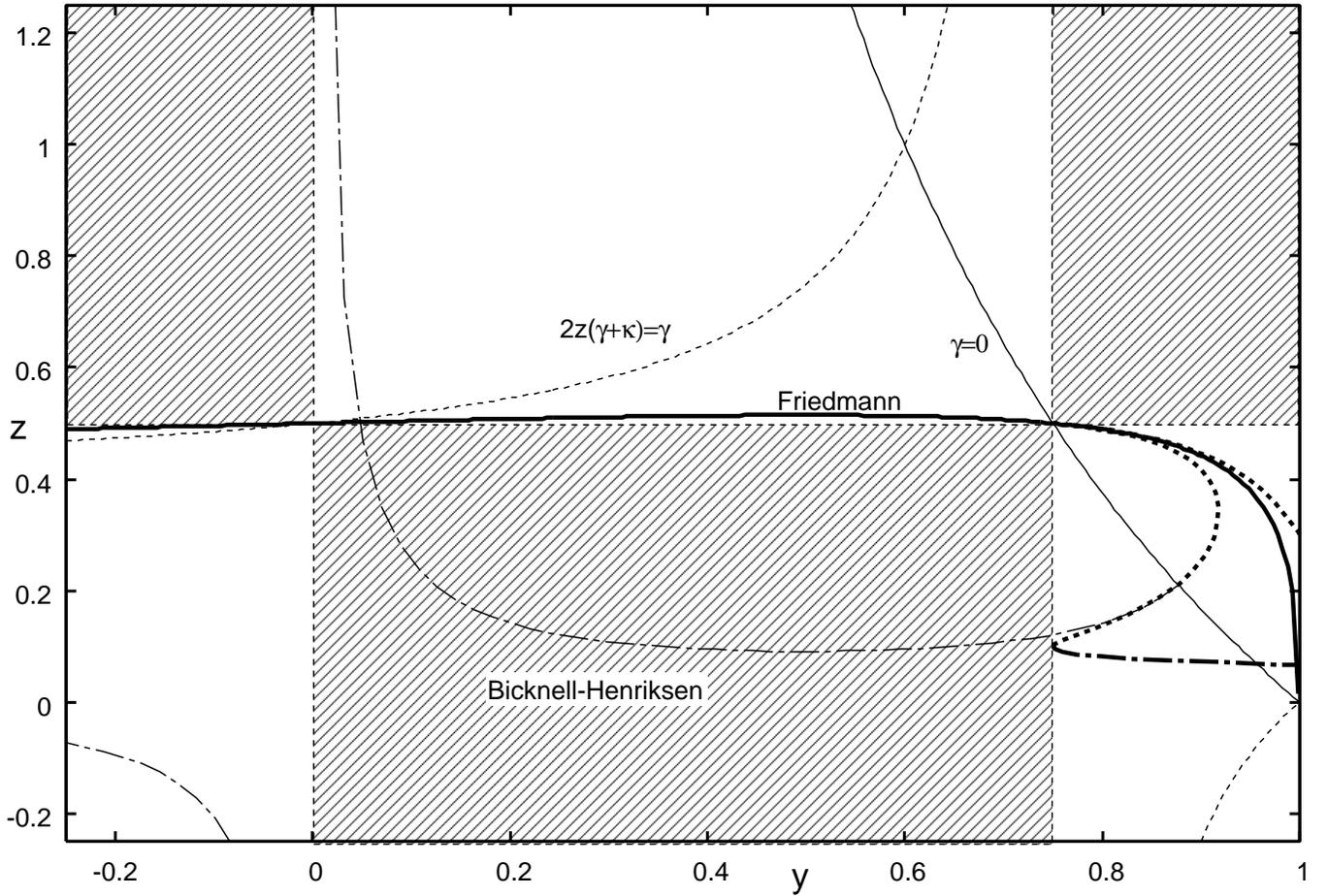}
\caption{ \label{fig:yz}
The solutions in the $yz$ plane for $4\pi\kappa^{2}=1/3$. 
The shaded regions are prohibited for a real scalar field. 
The expanding flat Friedmann solution is plotted
as a thick solid line,  
the curve $\gamma=0$ as a thin solid line, and 
the curve $2z(\gamma+\kappa)=\gamma$ as a thin
dashed line.
The gradient of the scalar field 
can change from timelike to spacelike
on the last two curves and also on $y=0$.
Note that the parts of the $2z(\gamma+\kappa)=\gamma$
and $\gamma=0$ curves in $z>1/2$
correspond to the null conditions for different solution branches 
but are shown together for convenience.
The flat Friedmann solution has a cosmological apparent horizon at $(y=0, z=1/2)$, a particle horizon at $(y=3/4, z=1/2)$ and ends at
 $(y=1, z=0)$.
Two numerical solutions within the particle horizon are 
shown by the thick dotted lines.
The upper one is a positive branch 
and goes directly
to the negative mass region ($y>1$).
The lower one reaches $y=3/4$, where it is converted from the positive branch to the negative branch (shown by the thick dashed-dotted
 line) before going to the negative mass region.
The Bicknell-Henriksen null-dust black hole
solution, which connects to the lower numerical solution, 
is indicated by a thin dashed-dotted line, 
but this is never realized by a scalar field. The black hole has an event horizon where the solution intersects $z=1/2$ and an apparent horizon at $(y=0, z =\infty).$}
\end{figure*}

\subsection{Brady's formulation}
Let us now return to the original scalar field description, in which the stress-energy tensor is
given by Eq.~(\ref{scalarstress}).
If we adopt (non-comoving) Bondi coordinates ($v,r$), where $v$ is advanced time, the metric can be written as
\begin{equation}
ds^{2}=-g\bar{g}dv^{2}+2g dv dr+r^{2}(d\theta^{2}
+\sin^{2}\theta d\phi^{2}),
\label{Bondimetric}
\end{equation}
where $g$ and $\bar g$ are functions of $v$ and $r$. This applies 
whether the gradient of the scalar field is timelike, 
null or spacelike. 
In this coordinate system, for a self-similar solution we can take the independent variable to be $\xi \equiv \ln (r/|v|)$ and then define the following variables:
\begin{eqnarray}
&& y(\xi)=\frac{\bar{g}}{g}=1-\frac{2m}{r}, \quad
z(\xi)=
\bar{g}^{-1} x,
\nonumber \\
&& \phi=\bar{h}(\xi)-\kappa \ln |v|, \quad  \gamma(\xi)=\dot{\bar{h}}(\xi),
\label{variables}
\end{eqnarray}
where $x= r/v = \pm e^{\xi} $ has the same sign as $v$ and a dot 
denotes a derivative with respect to $\xi$.
Note that $\phi$ depends on $\xi$  in this case, because we are not using
comoving coordinates. The quantity $z$ will play an analogous role to
the quantity $V$ in the stiff fluid picture. The quantity $\gamma$ is
related to the gradient of the scalar field; the field itself never
appears explicitly in the equations.  

With the self-similarity ansatz, the Einstein equations and equations of
motion for the scalar field reduce to the following ODEs:
\begin{eqnarray}
(\ln g)^{\cdot} &=&4\pi {\dot{\bar h}}^{2},\\
 {\dot {\bar g}}&=&g-{\bar g},\\
g \left(\frac{{\bar g}}{g}\right)^{\cdot}
&=&8\pi ({\dot {\bar h}}+\kappa) [x({\dot {\bar h}}+\kappa)-{\bar g}{\dot {\bar h}}],\\
 ({\bar g}{\dot {\bar h}})^{\cdot}+{\bar g}{\dot {\bar h}}&=&
2x\left({\dot {\bar h}}+{\ddot {\bar h}}+\kappa\right).
\end{eqnarray}
We then have an autonomous system~\cite{brady1995}:
\begin{eqnarray}
\dot{z}&=&z(2-y^{-1}), 
\label{eq:zdot}\\
\dot{y}&=&1-(4\pi \gamma^{2}+1)y, 
\label{eq:ydot} \\
(1-2z)\dot{\gamma}&=&2\kappa z-\gamma(y^{-1}-2z),
\label{eq:gammadot}
\end{eqnarray}
with the constraint equation
\begin{equation}
y[(1+4\pi\kappa^{2})-4\pi (\gamma+\kappa)^{2}(1-2z)]=1.
\label{eq:const2}
\end{equation}
Note from Eq.~(\ref{variables}) that $y < 1$ if the Misner-Sharp mass is positive.
This system was investigated 
in the context of cosmic censorship in~\cite{christodoulou1986}
and, for $\kappa=0$, one has the exact Roberts solution~\cite{roberts1989}.
Eq.~(\ref{eq:const2}) gives two values of $\gamma$ for each $(y,z)$:
\begin{equation}
\gamma=-\kappa\pm\sqrt{\frac{1+4\pi\kappa^{2}-y^{-1}}{4\pi(1-2z)}}\, ,
\label{eq:branch}
\end{equation}
with the positive (negative) branch corresponding to positive
(negative) values of $\gamma+\kappa$.
The structure of this system of equations is the same as in the stiff fluid description. 
Since Eq.~(\ref{eq:branch}) gives $\gamma$ in terms of $y$ and $z$, once we have determined 
initial values satisfying the constraint (\ref{eq:const2}),
Eqs.~(\ref{eq:zdot}) and (\ref{eq:ydot}) together with Eq.~(\ref{eq:branch})
form a two-dimensional dynamical system and Eq.~(\ref{eq:gammadot})
is evolved freely.

Special physical significance is
associated with the gradient of the scalar field being null. 
This is because the scalar field can no longer be identified with a stiff
fluid
once the gradient has turned from timelike to spacelike.
Since this gradient is
\begin{equation}
\phi^{,c}\phi_{,c} = -
\gamma y [2z(\gamma+\kappa)-\gamma]r^{-2},
\label{null}
\end{equation}
it can become null under three conditions: $y=0$, $\gamma=0$ and $2z(\gamma+\kappa)=\gamma$.
From Eq.~(\ref{eq:const2}), the condition $\gamma=0$ gives
\begin{equation}
z=\frac{1-y}{8\pi \kappa^{2}y}\, ,
\label{eq:gamma0}
\end{equation}
which always corresponds to the positive branch of Eq.~(\ref{eq:branch}). From Eq.~(\ref{eq:const2}), the condition
$2z(\gamma+\kappa)=\gamma$ gives
\begin{equation}
z=\frac{1-y}{2[1-(1+4\pi\kappa^{2})y]} \, .
\end{equation}
Since $\gamma+\kappa=\kappa/(1-2z)$, 
this corresponds to the positive (negative) branch 
of Eq.~(\ref{eq:branch}) for $z < (>) 1/2$. 
These three conditions for the gradient of the scalar field to be null 
are shown in Fig.~\ref{fig:yz} for $\kappa=1/\sqrt{12\pi}$. Note that the positive and negative branches of Eq.~(\ref{eq:branch}) should not really be represented in
the same $yz$ diagram, since the conditions $\gamma=0$ and $2z(\gamma+\kappa)=\gamma$ clearly cannot be satisfied simultaneously, but we do so for convenience. 

If the gradient is spacelike, the energy density $\epsilon$ given by Eq.~(\ref{eq:epsilon}) is negative.
This is not unphysical (eg. the weak energy condition is always
satisfied), even though the identification 
with a stiff fluid is no longer possible.
However, $\gamma$ must be real and, from Eq.~(\ref{eq:const2}), this applies within three regions in the $yz$ plane:
\[ z>\frac{1}{2} \quad\mbox{and}\quad  0<y<\frac{1}{1+4\pi \kappa^{2}}, \]
or 
\[ z<\frac{1}{2} \quad\mbox{and}\quad  y>\frac{1}{1+4\pi \kappa^{2}}, \]
or 
\[
  z<\frac{1}{2} \quad\mbox{and}\quad y<0.\]
These are the three unshaded regions in Fig.~\ref{fig:yz}.
Although one might speculate that the complex $\gamma$ region could be associated with a 
complex scalar field, this cannot be confirmed within the
context of the present single-field analysis.

\subsection{Similarity and trapping horizons}
\label{subsec:similarity_trapping_horizons}
From Eqs.~(\ref{Bondimetric}) and (\ref{variables}), the induced metric on the surface $\xi=\mbox{const}$ is 
\begin{equation}
ds^{2}= -y^{-1}z^{-2}(1-2z)dr^{2}+r^{2}(d\theta^{2}+\sin^{2}\theta d\phi^{2}).
\label{eq:scalarmetric}
\end{equation}
Thus, providing $y\ne 0$, $z=1/2$ corresponds to a similarity horizon and this
is either a particle horizon or an event horizon.
Although $z=1/2$ plays a similar role in the field description
to $V^{2}=1$ in the 
fluid description, we will show that the degeneracy
in the fluid description is resolved in the field description.
From Eq.~(\ref{eq:const2}), $\gamma$ can be finite at $z=1/2$ only if $y=1/(1+4\pi\kappa^{2})$.

If $y=0$, one has a trapping horizon. This notion was defined in~\cite{hayward1993,hayward1996}
and has turned out to be a very useful generalization of the concept  of 
apparent horizons.
Past trapping horizons include cosmological apparent horizons,
while future ones include black hole apparent horizons. 
Equation (\ref{eq:const2}) implies that either $z$ or $\gamma$ must diverge when $y=0$. 
From Eqs.~(\ref{Bondimetric}) and (\ref{variables}), $dr/dv=\bar{g}/2=e^{\xi}/(2z)$ 
is an outgoing null surface, so
a future trapping horizon has infinite $|z|$ and finite $\gamma$, while
a past trapping horizon has $z=1/2$ and infinite $\gamma$. 
More precisely, 
$y \propto z^{-1}$ in the vicinity of future trapping horizons, while 
$y \propto (1-2z)$ and $\gamma\propto y^{-1}$ in the vicinity of past trapping horizons, where we have used Eq.~(\ref{eq:const2}). 
In the latter case, the apparent singularity arises from the fact
that $\xi$ or $r$ has an extremum along ingoing null rays $v=\mbox{const}$
on past trapping horizons. 

From these properties of similarity and trapping horizons, 
we can deduce the following.
A similarity horizon with $z=1/2$ cannot be a future 
trapping horizon. $z=1/2$ is satisfied
at both similarity and past trapping horizons.
We can show that a black hole event horizon is untrapped with 
negative ingoing null expansion and positive outgoing null expansion,
where we need to assume the existence of future null infinity
for the definition of a black hole event horizon.
This is because, if both expansions were negative, 
no causal curve in the past neighbourhood of 
the black hole event horizon could 
reach future null infinity. On the other hand, if both expansions were positive,
there would exist a causal curve in the future neighbourhood of the 
black hole event horizon 
which could reach future null infinity.
It also follows that 
a black hole event horizon cannot coincide with a past trapping horizon.
Therefore, a black hole event horizon is untrapped, i.e. 
its null expansions have different signs.

\subsection{Transformation between the two
formulations}

The key step is to obtain the correspondence between two
equivalent expressions for the three 
quantities $2m/r$, $8\pi \epsilon r^{2}$ and $V$. 
The first two can be written as
\begin{eqnarray}
2m/r&=&AE |V|^{-1}=1-y, \label{eq:AEVinv}\\
8\pi \epsilon r^{2}&=&E |V|^{-1} = 4\pi 
\gamma y [2z(\gamma+\kappa)-\gamma].
\label{eq:eta_gamma}
\end{eqnarray}
To obtain the expression for $V$ in the field description, we
introduce the following vector fields:
\begin{equation}
\bar{u}_{a}=\phi_{,a}, \quad \bar{n}_{a}=\epsilon_{a}^{~b}\phi_{,b},
\end{equation}
where $\epsilon_{ab}$ is the totally antisymmetric tensor, which 
satisfies $\epsilon_{a}^{~c}\epsilon_{cb}=-g_{ab}$.
Since these two vectors satisfy the normalisation relation
\begin{equation}
\bar{u}_{a}\bar{u}^{a}=-\bar{n}_{b}\bar{n}^{b},
\end{equation}
the velocity function $V$ can be written as
\begin{equation}
V=-\frac{\bar{n}^{a}\xi_{a}}{\bar{u}^{a}\xi_{a}}
\end{equation}
if $\phi_{,a}$ is timelike; here $\xi^{a} \equiv (\partial_{\xi})^{a}$.
Since $\xi$ is a function of $X$,
this definition applies for any self-similar spacetime with
a scalar field.
Calculating the right-hand side of this equation 
in the Bondi coordinates and assuming $\kappa\ne 0$, we have
\begin{equation}
V= 1+\left(2-\frac{1}{z}\right)\frac{\gamma}{\kappa},
\label{eq:V2_zgamma}
\end{equation}
where $\epsilon^{01}$ is chosen to be positive. 
These relations give the explicit coordinate transformations
between the fluid and field descriptions if $\phi_{,a}$ is timelike.

Equation~(\ref{eq:V2_zgamma}) means that the condition $V^{2}=1$ in the fluid description 
separates into two independent 
conditions, $z=1/2$ and $\gamma=0$, in the field description.
From Eq.~(\ref{eq:scalarmetric}), the condition $z=1/2$ implies that 
the $\xi=\mbox{const}$ surface 
is null unless $y=0$. On the other hand, from Eq.~(\ref{null}),
the condition $\gamma=0$ implies $\phi_{,a}$ is null and this is plotted in Fig.~\ref{fig:yz} 
using Eq.~(\ref{eq:gamma0}).
There is no coordinate singularity even when $\phi_{,a}$ is null
in the field description, and this is very important 
for the present analysis.
As $\gamma$ approaches zero, 
Eqs.~(\ref{eq:e2nu}), (\ref{eq:eta_gamma}) and (\ref{eq:V2_zgamma}) imply that 
$1-V^{2}$ is proportional to $\gamma$ and that $e^{2\nu}(1-V^{2})$ has a finite limit 
$e^{2X}(1-2z)/(yz^{2})$. This 
 is zero only if $z=1/2$, so Eq.~(\ref{eq:xi_const_surface}) implies that 
the similarity surface is null only in this case. 

\subsection{Friedmann solution}
We now consider the flat Friedmann solution. This has $A=1/3$ everywhere in the fluid
description and a regular similarity horizon. Since $4\pi\kappa^{2}=1/3$,
Eqs.~(\ref{eq:V2'})--(\ref{eq:eta'}) imply 
\begin{equation}
V^{2}=C^{2}e^{4X/3},\quad A=\frac{1}{3},\quad E=\frac{3}{4}C^{3}e^{2X}.
\label{eq:frw_comoving}
\end{equation}
The integration constant $C$ comes from 
the remaining rescaling freedom of $t$.
Using Eqs.~(\ref{eq:const2}), (\ref{eq:AEVinv}), (\ref{eq:eta_gamma}),
(\ref{eq:V2_zgamma}) and (\ref{eq:frw_comoving}), it can be shown that
the Friedmann solution is described by:
\begin{eqnarray}
z&=&3\displaystyle{\frac{\sqrt{1-y}(1-\sqrt{1-y})}{y(1+2\sqrt{1-y})}} 
\quad \mbox{(expansion)}, \label{eq:expand_frw} \\
z&=&\displaystyle{\frac{\sqrt{1-y}(1+\sqrt{1-y})}{y(1+2\sqrt{1-y})}}
\quad \mbox{(collapse).} \label{eq:collapse_frw}
\end{eqnarray}
The discussion about trapping horizons confirms that
the first and second solutions represent the expanding and 
collapsing Friedmann solutions, respectively. 
The expanding solution is shown by the thick solid line in Fig.~\ref{fig:yz} and this crosses $z=1/2$ 
at $y=0$ and $y=3/4$. From Eq.~(\ref{eq:V2_zgamma}) $V=1$ at $y=3/4$ and so this corresponds to the cosmological particle horizon.
Eq.~(\ref{eq:expand_frw}) implies 
$y^{-1}(1-2z)\to -1/12$ as $y\to 0$, which leads to $V=2$ from Eqs.~(\ref{eq:branch}) and (\ref{eq:V2_zgamma}).
Therefore this surface is spacelike 
and corresponds to a past trapping horizon.
This is the cosmological apparent horizon,
which is outside the particle horizon and 
coincides with
the Hubble horizon~\cite{hc2005a}. This illustrates that the condition
$z=1/2$, like the condition $V=1$, does not always correspond to a null
surface. 
The causal structure of the Friedmann solution in this case is shown
in Appendix~\ref{sec:friedmann}.

\subsection{Structure of similarity horizons}
\label{subsec:structure_similarity_horizon}
As we have seen, a similarity horizon is defined as a null
similarity surface. It is denoted by $\xi=\xi_{\rm s}$ and characterized by $z=1/2$ and $y\ne 0$.
Since this corresponds to a singular point of the autonomous system 
(\ref{eq:zdot})--(\ref{eq:gammadot}), the usual uniqueness of 
solutions may break down there. 
The behaviour of solutions around $\xi=\xi_{\rm s}$ can be analysed by dynamical 
systems theory~\cite{bh1978b,harada2001,hm2003}.
First, we restrict our attention to $\kappa\ne 0$
and assume that $\gamma$ has a finite limit at $\xi=\xi_{\rm s}$. 
Then $(1-2z)\dot{\gamma}$ tends to zero at
the similarity horizon from Eq.~(\ref{eq:gammadot}).
Otherwise, $\gamma \propto \ln |\xi-\xi_{\rm s}|$ 
in the vicinity of $\xi=\xi_{\rm s}$ and this necessarily 
implies a spacetime singularity unless $y=0$.

The full details of the analysis are described in Appendix~\ref{sec:similarity_horizon_massless}. Here we describe the important qualitative results. The regularity of the similarity horizon requires
\begin{equation}
y_{\rm s}=\frac{1}{1+4\pi \kappa^{2}}, \quad
\gamma_{\rm s}=\frac{1}{4\pi\kappa},
\label{ygamma}
\end{equation}
from Eqs.~(\ref{eq:gammadot}) and (\ref{eq:const2}).
We linearise the ODEs around the similarity horizon and 
represent the solutions as trajectories in $(z,y,\gamma, \xi)$ space. Generically regular solutions can only cross the similarity surface along two directions in this space, corresponding to two
eigenvectors ${\bf e}_{1}$ and ${\bf e}_{2}$. The associated values of 
$\dot{\gamma}_{\rm s}$ are:
\begin{equation}
 \dot{\gamma}_{1}=\kappa 
\frac{(1-\alpha)(1+\alpha)}{\alpha^{2}(1-2\alpha)},
\quad \dot{\gamma}_{2}=\pm \infty
\label{gammadot}
\end{equation}
where $\alpha\equiv 4\pi\kappa^{2}$. 
Along the second eigenvector, there is a solution which describes
a density discontinuity surface at fixed $\xi=\xi_{\rm s}$. 
We call this a shock-wave solution.

As discussed in Appendix~\ref{sec:similarity_horizon_massless}, the signs of the eigenvalues associated with the two eigenvectors determine the qualitative behaviour of 
solutions around the equilibrium point.
For $0<4\pi \kappa^{2}<1/2$, 
the similarity horizon is a non-degenerate node, with ${\bf e}_{1}$
and ${\bf e}_{2}$ corresponding to the secondary and primary directions, respectively. 
This means that there is a one-parameter family of 
solutions which belong to the ${\bf e}_{2}$ direction and 
an isolated solution which belongs to the ${\bf e}_{1}$ direction.
In particular, this includes the case with $4\pi\kappa^{2}=1/3$.
Since the flat Friedmann solution is along ${\bf e}_{1}$,
this is isolated, while all other solutions have a sound-wave with diverging $\dot{\gamma}$
at the similarity horizon. One of them is a shock-wave.
For $4\pi \kappa^{2}=1/2$, 
the similarity horizon is a degenerate node.
For $1/2<4\pi \kappa^{2}<1$, 
it again becomes a non-degenerate node but with ${\bf e}_{1}$
and ${\bf e}_{2}$ corresponding to the primary and secondary directions, respectively. 
For $4\pi \kappa^{2}>1$, 
the similarity horizon is a saddle point. 
This means that there are two isolated 
solutions associated with ${\bf e}_{1}$ and ${\bf e}_{2}$.

Brady~\cite{brady1995} has analysed the structure of the system 
governed by Eqs.~(\ref{eq:zdot})--(\ref{eq:const2}) for all values of the constant $\kappa$. For $0<4\pi\kappa^{2}<1$,  which includes our choice of $4\pi\kappa^{2}=1/3$, the allowed region 
in the $yz$ plane  is 
similar to that shown in Fig.~\ref{fig:yz}.
The solution can cross $z=1/2$ with finite $\gamma$
only if $y_{\rm s}$ and $\gamma_{\rm s}$ are given by Eq.~(\ref{ygamma}).
Brady has also made a non-linear analysis of 
the similarity horizon and we expand his analysis in Appendix~\ref{sec:follow_brady},
correcting a typographical error.
A comparison of the two analyses shows that the qualitative behaviour obtained
by linearizing the ODEs reflects the true non-linear behaviour.

\subsection{Non-existence of self-similar PBHs with a massless scalar field}
If the initial 
perturbation extends to infinity, so that the specific binding
energy is negative there, the perturbation never goes to zero
because the angular part of the metric acquires a numerical 
factor which results in a solid angle deficit~\cite{mkm2002}. 
Although  these solutions are sometimes described as asymptotically Friedmann~\cite{ch1974}, they could in principle be observationally distinguished  from the exact Friedmann solution by the angular diameter test. However, inflation itself would not produce models of this kind. 
We therefore first seek a black hole solution surrounded by an exact flat Friedmann
solution beyond the 
sonic point, which is also the particle horizon, and consider matching to other interior self-similar solutions there. 

\begin{figure}[htbp]
\includegraphics[scale=0.8]{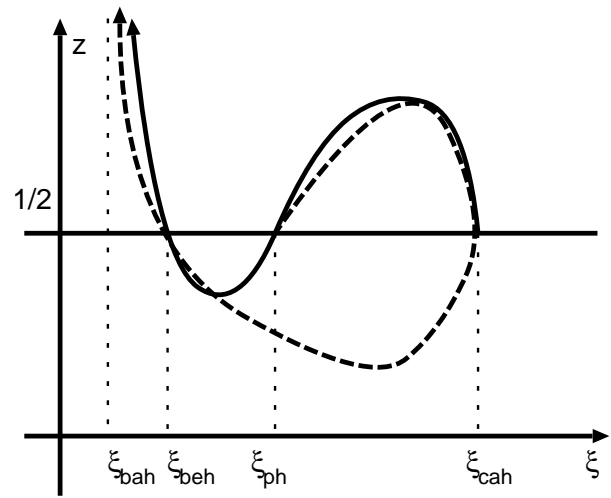}
\caption{\label{fig:expected_z} The required form of $z(\xi)$
for a black hole solution is plotted as a solid line if 
the cosmological apparent horizon is spacelike and 
as a dashed line if it is timelike.}
\end{figure}

The required behaviour of the function $z$ 
for a black hole is shown by the solid line 
in Fig.~\ref{fig:expected_z}. 
As we move along the ingoing null surface $v=\mbox{const}$ from the cosmological
apparent horizon, $\xi$ decreases from $\xi_{\rm cah}$.
Thus
$z(\xi)$ increases from $1/2$, reaches a maximum, 
turns to decrease
and then crosses 1/2 at $\xi_{\rm ph}$. Up to this point, the solution 
is described by the Friedmann solution, 
but the uniqueness of the solution breaks down at $\xi_{\rm ph}$. Because the Friedmann solution approaches the similarity horizon along a secondary direction, it has to connect to a member of the one-parameter family of solutions associated with the primary direction. As we decrease $\xi$ further,  in order to 
have a black hole solution,
$z$ should first decrease, reach a minimum,
begin to increase and then cross 1/2 at 
the black hole event horizon $\xi=\xi_{\rm beh}$. 
It should then diverge at the black hole apparent horizon 
$\xi=\xi_{\rm ah}<\xi _{\rm beh}$ (which is necessarily within the event horizon), 
although this feature is not required for the following discussion.

It is easy to see that the required behaviour is impossible. Because
$z<1/2$ is required for $\xi_{\rm beh}<\xi<\xi_{\rm ph}$, we need 
$y>3/4$ to avoid the shaded region in Fig.~\ref{fig:yz} and Eq.~(\ref{eq:zdot}) then implies $2/3<\dot{z}/z<2$. 
However, this means that $z$ cannot have an extremum, which gives a contradiction.
The solution therefore either enters
the negative mass region beyond $y=1$
or approaches $z=0$ as $\xi \to -\infty$ (i.e. as $r \to 0$), corresponding to the exact Friedmann solution.
The only way to invalidate this proof would be to find a solution which goes through the lower shaded region in Fig.~\ref{fig:yz}. One can never reach the black hole event horizon by going around this region. 

Figure~\ref{fig:yz} shows solutions obtained 
through numerical integration of Eqs.~(\ref{eq:zdot}) -- 
(\ref{eq:gammadot}) inside the particle horizon.
Some solutions directly cross $y=1$ and go into the negative 
mass region with $y>1$. However, there are also solutions 
which reach $y=3/4$ at $z<1/2$. 
These necessarily cross the curve $\gamma=0$, which
means that they have the required behaviour for $V^{2}$
in Fig.~\ref{fig:expected_V2} but not for $z$
in Fig.~\ref{fig:expected_z}. 
One could also consider solutions which pass smoothly into the region
where the scalar field gradient is spacelike. However, in this case, the above
proof still applies because $z$ remains monotonic, so there is again no
black hole solution.  
Such solutions have to touch the line
$y=3/4$, after which they become negative-branch solutions
and eventually enter the negative mass region with $y>1$.

Although we have assumed that the spacetime exterior to the 
black hole solution is an exact Friedmann solution, the proof also applies 
if it is asymptotically 
Friedmann. 
For the equality $4\pi \kappa^{2}=1/3$ must still be satisfied in order
to allow the exact Friedmann solution, the black hole solution must still
have a particle horizon at $z=1/2$ and there must still 
be a region where $0<z<1/2$ inside the particle horizon. Therefore 
one can still conclude that 
$z$ cannot cross $z=1/2$ again from below. This proves that there is no 
black hole event horizon inside the particle horizon even for 
asymptotically Friedmann solutions. 

It should be noted that the situation is quite different for 
a perfect fluid with equation of state $p=k\rho$ ($0\le k<1$). 
In fact, in this case, analytic and numerical studies show that there {\it do} exist asymptotically
Friedmann self-similar 
solutions with
a particle horizon and black hole event horizon~\cite{ch1974,bh1978b,carryahil1990}.
The crucial point is that there is no extremum in $z$ as a function of $\xi$
in a subsonic region but there can be in a supersonic but subluminal
 ($z<1/2$) region for $0\le k<1$.
The reason why the non-existence proof works in the scalar case 
is that the sonic horizon in the scalar field system
coincides with the similarity horizon. Since $z$ cannot have an 
extremum in a subluminal region, the non-existence of 
self-similar black holes in the scalar field system follows immediately.

If there is a shock-wave, it can be located only at the 
similarity horizon. As we have seen, such a solution
has $z=1/2$ and $y=(1+4\pi\kappa^{2})^{-1}$ but changing $\gamma$.
If we consider a solution which is interior to this shock-wave 
and assume a regular junction, it must also
have $z=1/2$ and $y=(1+4\pi\kappa^{2})^{-1}$ there. Because of regularity, $\gamma$  must have the
same value $(4\pi\kappa)^{-1}$ at the similarity horizon 
and in the interior solution. In fact, this shows that
no regular shock-wave is allowed.
We can still have a sound-wave solution 
with diverging $\dot{\gamma}$ for $0<4\pi\kappa^{2}<1/2$,
which is included in the above proof.
If we do not assume regularity of the similarity horizon,
we may consider an extension beyond the 
similarity horizon to a solution with a different $\kappa$. 
However, the different value of $\kappa$ 
means a different value of $y$ for $z=1/2$ 
because $y=(1+4\pi\kappa^{2})^{-1}$ there.
One needs  
a singular null hypersurface because
of the discontinuity in the Misner-Sharp mass. 
We do not consider this possibility 
further here.

\subsection{Bicknell and Henriksen's PBH construction with a null dust}
\label{subsec:Bicknell_Henriksen_solution}
Bicknell and Henriksen~\cite{bh1978a} 
found an extension of the self-similar 
stiff fluid solution beyond the timelike 
$V^{2}=1$ surface, containing null dust.
The extension is given by an ingoing Vaidya solution~\cite{vaidya1951}, 
in which the line element is given by 
\begin{equation}
ds^{2}=- \left(1-\frac{2m(v)}{r}\right)dv^{2}+2dvdr+r^{2}(d\theta^{2}
+\sin^{2}\theta d\phi^{2}).
\label{eq:vaidya_metric}
\end{equation}
The stress-energy tensor for null dust can be written in the form
\begin{equation}
T_{ab}=\sigma l_{a}l_{b},
\label{eq:null_dust}
\end{equation}
where $l_{a}$ is a null vector.
For the Vaidya solution, $l_{a}=-\delta^{0}_{a}$ and 
$\sigma$ is written as
\begin{equation}
\sigma =\frac{1}{4\pi r^{2}}\frac{dm}{dv}.
\end{equation}
The similarity assumption implies
\begin{equation}
m(v)=\mu v,
\end{equation}
where $\mu$ is an arbitrary constant. Then, in terms of $y$, $z$
and $\xi$, Eqs.~(\ref{variables}) and (\ref{eq:AEVinv}) imply 
that this solution is
\begin{eqnarray}
y&=&1-2 \mu e^{-\xi}, \\
z&=&\frac{e^{\xi}}{1-2\mu e^{-\xi}}.
\end{eqnarray}
Eliminating $\xi$, we have
\begin{equation}
z=\frac{2\mu}{y(1-y)}.
\label{bhsolution}
\end{equation}
The junction between the Vaidya metric and the stiff fluid solution 
can be implemented on the line $\gamma=0$, given by Eq.~(\ref{eq:gamma0}).
For $4\pi\kappa^{2}=1/3$, we then have 
\begin{equation}
\mu=\frac{3}{4}(1-y_{*})^{2},
\end{equation}
where $y=y_{*}$ on the junction surface. Thus $0<\mu<3/64$ 
for $3/4<y_{*}<1$. We can see explicitly 
from Eqs.~(\ref{eq:zdot}) and (\ref{eq:ydot}) that 
$\dot{z}$ and $\dot{y}$ are continuous on this surface, which 
guarantees the continuity of the first and second 
fundamental forms of the junction surface.
For the extended solution to have a black hole event horizon, 
we require that 
$z$ go below $1/2$ for $y>0$ and Eq.~(\ref{bhsolution}) shows that this condition is necessarily satisfied. 
An example of an extended solution is shown
by a thin dashed-dotted line in Fig.~\ref{fig:yz}. 

If the massless scalar field $\phi$ has a null gradient,
the stress-energy tensor (\ref{scalarstress}) is equivalent to that 
for the null dust (\ref{eq:null_dust}), with 
$l_{a}=\sigma^{-1/2}\nabla_{a}\phi$. For this reason,
one might expect that this construction could be realised consistently
with a massless scalar field alone. Nevertheless, 
this is not true. If a null dust is equivalent to a massless scalar
field, the vorticity-free condition must be satisfied:
\begin{equation}
\nabla_{[a}(\sqrt{\sigma} l_{b ]})=0.
\label{eq:vorticity_free}
\end{equation}
However, for the Vaidya solution, we can see 
\begin{equation}
\nabla_{[0}(\sqrt{\sigma} l_{1]})=-\frac{\sqrt{\sigma}}{r}.
\end{equation}
This means that the vorticity-free condition is not satisfied 
for the Vaidya solution unless it is vacuum.
Therefore, the null dust which appears in the Vaidya metric
cannot be realized by a massless scalar field. 

The stiff fluid or massless scalar field in 
this construction turns into null dust on the timelike surface 
where $\gamma =0$, but it is then no longer describable as a 
massless scalar field. The Bicknell and Henriksen PBH solution 
exploits this feature and that is why this solution goes through 
the prohibited region where $\gamma$ is complex and why it 
circumvents the above proof that $z$ cannot have a minimum.
However, this may be regarded as physically contrived.

\section{Scalar field with a potential}

\subsection{Autonomous system}
When the scalar field $\phi$ has a potential $V(\phi)$, the existence of
self-similar solutions requires it to be of exponential form~\cite{we1997}:
\begin{equation}
V(\phi)=V_{0}e^{\sqrt{8\pi}\lambda\phi}.
\label{eq:potential}
\end{equation}  
Then the stress-energy tensor of the scalar field is 
\begin{eqnarray}
T_{ab}=\phi_{,a}\phi_{,b}-g_{ab}\left(\frac12\phi_{,c}\phi^{,c}+V_0e^{\sqrt{8\pi}\lambda\phi}\right).
\end{eqnarray}
With the self-similarity ansatz, the Einstein equations and the equations of motion for the scalar field reduce to the following ODEs:
\begin{eqnarray}
(\ln g)^{\cdot} &=&4\pi {\dot{\bar h}}^{2},\label{basic1}\\
{\dot {\bar g}}&=&g\left(1-8\pi x^2V_{0}e^{\sqrt{8\pi} \lambda{\bar h}}\right)-{\bar g},\label{basic2}\\
g \left(\frac{{\bar g}}{g}\right)^{\cdot}
&=&8\pi ({\dot {\bar h}}+\kappa) [x({\dot {\bar h}}+\kappa)-{\bar g}{\dot {\bar h}}],\label{basic3}\\
 ({\bar g}{\dot {\bar h}})^{\cdot}+{\bar g}{\dot {\bar h}}&=&\sqrt{8\pi}\lambda gx^2V_{0}e^{\sqrt{8\pi} \lambda{\bar h}}\nonumber \\
&&+2x\left({\dot {\bar h}}+{\ddot {\bar h}}+\kappa\right),\label{basic4}
\end{eqnarray}
where we use the same notations as for the massless case, except that
one requires $\kappa = (\sqrt{2\pi}\lambda)^{-1}$ 
in the presence of a potential.
This system has been investigated in the context of the kink instability 
analysis~\cite{mh2005a}. 

We define the functions $y$, $z$ and $\gamma$ in the same way as before [cf. Eq.~(\ref{variables})] and introduce a new function
\begin{equation} 
\beta \equiv 8\pi V_0\exp\left(2\xi + \frac{2{\bar h}}{\kappa}\right).
\label{eq:beta}
\end{equation}
Eqs.~(\ref{eq:zdot})--(\ref{eq:const2}) are then replaced by
\begin{eqnarray}
\dot{z}&=&z[2-y^{-1}(1-\beta)], \label{eq:zdot_pot}\\
\dot{y}&=&1-\beta-(4\pi \gamma^{2}+1)y, \label{ydot} \\
(1-2z)\dot{\gamma}&=&2\kappa z-\gamma[y^{-1}(1-\beta)-2z]+\frac{\beta}{4\pi
 \kappa y},\label{eq:gammadot'}\\
\dot{\beta}&=&2 \beta\left(1+\frac{\gamma}{\kappa}\right)
\end{eqnarray}
with the constraint equation
\begin{equation}
y[(1+4\pi\kappa^{2})-4\pi (\gamma+\kappa)^{2}(1-2z)]=1-\beta.\label{eq:const2'}
\end{equation}
The solutions are therefore constrained to lie on a 3-dimensional hypersurface
in a 4-dimensional phase space.
Because the dimension of the phase space is 
higher than that for the massless case,
the analysis becomes more complicated.
However, the number of degrees of freedom for the self-similar 
solutions is the same
because the parameter $\kappa$ is fixed by the potential,
whereas it is freely chosen for the massless case.

\subsection{Friedmann solution}
The flat Friedmann solution satisfies these equations providing
$\kappa=(\sqrt{2\pi}\lambda)^{-1}$.
This is in contrast to the massless case,
in which the Friedmann solution requires $4\pi\kappa^{2}=1/3$. 
In the presence of a potential, 
the scale factor is given by 
\begin{equation}
a(t)\propto t^{4\pi\kappa^{2}}\propto t^{2/\lambda^{2}}.
\end{equation}
Thus the expansion 
is decelerated 
for $0<4\pi\kappa^{2}<1$, accelerated for $4\pi\kappa^{2}>1$
and with a constant velocity for $4\pi\kappa^{2}=1$.
We call the scalar field in the accelerated case a ``quintessence'' field.
As discussed in Appendix~\ref{sec:friedmann}, there are five 
types of causal structures,
corresponding to $0<4\pi\kappa^{2}<1/2$,
$4\pi\kappa^{2}=1/2$, $1/2<4\pi\kappa^{2}<1$,
$4\pi\kappa^{2}=1$ and $4\pi\kappa^{2}>1$. 

\subsection{Structure of similarity horizons}
As seen in 
Section~\ref{subsec:similarity_trapping_horizons}, 
similarity horizons are characterised by 
$z=1/2$ and $y\ne 0$, future trapping horizons 
by $y=0$ and $|z|=\infty$, and past trapping horizons
by $y=0$ and $z=1/2$.
From Eqs.~(\ref{eq:gammadot'}) and (\ref{eq:const2'}), the regularity of the similarity horizon yields
\begin{eqnarray}
y_{\rm s}=\frac{1-\beta_{\rm s}}{1+4\pi\kappa^2}, 
~\gamma_{\rm s}=\frac{\beta_{\rm s}+4\pi\kappa^{2}}{16\pi^{2} \kappa^{3} (1-\beta_{\rm s})},
\label{eq:regular_similarity_horizon_potential}
\end{eqnarray}
where we have assumed $\beta_{\rm s} \ne 1$.
Eq.~(\ref{eq:const2'}) shows that
$\beta_{\rm s}=1$ is only possible when $\gamma=\infty$, which means that
the similarity horizon
coincides with the past trapping horizon. Note that Eq.~(\ref{eq:regular_similarity_horizon_potential}) reduces to Eq.~(\ref{ygamma}) when $\beta_{\rm s}=0$.

As is before, we
linearise the ODEs around the similarity horizon, which again corresponds to a singular point in the presence of a potential. The situation is qualitatively similar to the massless case, except that the solutions are now represented as trajectories in $(z,y,\gamma, \xi, \beta)$ space.  Generically regular solutions can only cross the similarity surface along two 
eigenvectors ${\bf e_{1}}$
and ${\bf e_{2}}$. As before, these are associated with a finite and infinite value of $\dot{\gamma}_{\rm s}$, respectively. We do not give the analogue of Eq.~(\ref{gammadot}) explicitly, since it is very complicated, but the infinite value is associated with
a shock-wave solution, for which  
$y=y_{\rm s}$, $z=1/2$, $\beta=\beta_{\rm s}$, 
$\xi=\xi_{\rm s}$ but $\gamma$ changes.
This belongs to the one-parameter family of solutions which cross the 
similarity horizon along ${\bf e_{2}}$.
As before, the similarity horizon behaves as a node 
for $0<4\pi\kappa^{2}<1$ or as a saddle for $4\pi\kappa^{2}>1$. 
Appendix~\ref{sec:similarity_horizon_potential} provides
the details of the linearised ODEs and eigenvectors for 
this case.

\subsection{Non-existence of self-similar PBH in decelerating Friedmann universe}
We now consider solutions which are exactly 
Friedmann outside the particle horizon. 
The Friedmann background is decelerating for $0<4\pi\kappa^{2}<1$, i.e.
for a steep potential with $\lambda^{2}>2$.
As shown in Appendix~\ref{sec:friedmann},
the particle horizon in this case is a similarity horizon and the Penrose diagram for a solution containing a black hole would be as indicated in Fig.~\ref{fig:dec_pbh_penrose}. 
Because of the nodal 
nature of the similarity horizon, there is a one-parameter family of interior solutions which can 
be matched to this background. However, we now show that no 
interior self-similar 
solution can contain a black hole event horizon.

For $0<4\pi\kappa^{2}<1/2$, i.e. for $\lambda^{2}>4$, 
the cosmological apparent horizon 
is spacelike and outside the particle horizon, 
as shown in Appendix~\ref{sec:friedmann}.
So if we have a black hole event
horizon, as $\xi$ decreases monotonically
along a future-directed ingoing null ray from the particle horizon,
$z$ decreases from $1/2$, reaches a minimum, increases 
and then crosses $1/2$ again.
$y$ is positive in this region.
The required behaviour of $z$ is therefore as plotted with a solid curve in 
Fig.~\ref{fig:expected_z}. 
For $1/2 < 4\pi\kappa^{2}<1$, i.e. $2<\lambda^{2}<4$, 
the cosmological apparent horizon 
is timelike and inside the particle horizon, as shown in 
Appendix~\ref{sec:friedmann}.
Since a black hole event horizon
is untrapped, as $\xi$ decreases monotonically
along a future-directed ingoing null ray from the cosmological 
apparent horizon,
$z$ decreases from $1/2$, reaches a minimum, increases 
and then crosses $1/2$ again.
$y$ increases from zero as we move along
a future-directed ingoing null ray from the cosmological 
apparent horizon. 
The required behaviour of $z$ is plotted with a dashed curve in 
Fig.~\ref{fig:expected_z}. 
For $4\pi\kappa^{2}=1/2$, i.e. $\lambda^{2}=2$,
the cosmological apparent horizon 
coincides with the particle horizon, as shown in 
Appendix~\ref{sec:friedmann}.

We now show that such a solution cannot exist. 
From Eqs.~(\ref{eq:zdot_pot}) and (\ref{eq:const2'}), we have
\begin{equation}
\frac{\dot{z}}{z}=1-4\pi\kappa^{2}+4\pi(\gamma+\kappa)^{2}(1-2z).
\label{eq:z'potential}
\end{equation}
Therefore, if $4\pi\kappa^{2}<1$, 
$z$ cannot have an extremum for $z<1/2$. Hence there is 
no self-similar black hole solution in this case. 
The proof also applies for the asymptotic Friedmann background.
Thus the conclusion that there is no self-similar black hole solution in the case without a potential also applies for a decelerating universe.

\subsection{Non-existence of self-similar PBH in accelerating universe}
If $4\pi\kappa^{2}=1$, or $\lambda^{2}=2$, the Friedmann universe 
expands with a constant speed.
In this case, the big bang singularity is an outgoing null surface and 
future null infinity is ingoing null surface, as shown in 
Appendix~\ref{sec:friedmann}.
We proved in Appendix~\ref{sec:friedmann} that there is 
no similarity horizon in this case,
so we cannot match the exterior Friedmann 
background to an interior black hole solution. We therefore exclude this 
case.

\begin{figure}
\includegraphics[scale=1]{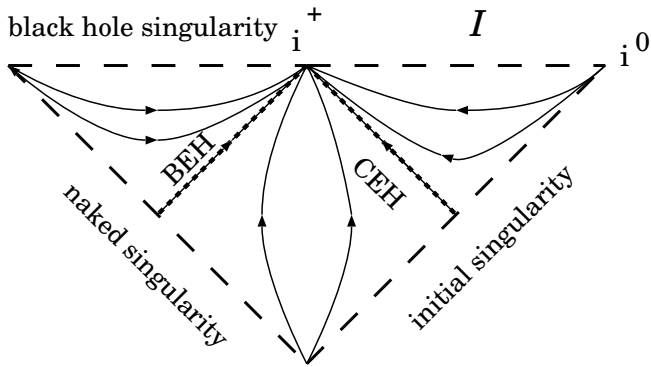}
\caption{\label{fig:acc_pbh_penrose}
The causal structure of self-similar black holes in an accelerating universe.
The trajectories of similarity surfaces are shown.
There is a null big bang singularity, a null naked singularity and
a spacelike black hole singularity.
The cosmological event horizon (CEH) and the black hole event horizon
(BEH) are both null similarity surfaces, i.e., similarity horizons.}
\end{figure}

For $4\pi\kappa^{2}>1$,
i.e. for a flat potential with $0<\lambda^{2}<2$,
the scalar field with a potential acts as dark energy,
which causes accelerated expansion.
The Penrose diagram for a self-similar black hole spacetime is now different 
from the decelerating case, as indicated in Fig.~\ref{fig:acc_pbh_penrose}. 
This reflects the fact that the causal structure of the flat Friedmann universe is itself modified, as discussed in Appendices~\ref{sec:friedmann} and \ref{sec:friedmann_self-similar}. 
The spacetime now has 
a null big bang singularity, as well as a null naked
singularity, a spacelike black hole singularity and black hole event horizon
inside the cosmological event horizon.
The similarity surfaces are spacelike outside the
cosmological event horizon and inside the black hole event horizon, timelike in between them
and null on the horizons themselves.
So the similarity horizons can be identified with the cosmological event
horizon or black hole event horizon.

In this case, the sign of the right-hand side of
Eq.~(\ref{eq:z'potential}) is apparently indefinite, so we cannot
immediately apply the earlier proof. However, the nature of the
similarity horizon in the Friedmann solution now plays a crucial
role. Because the cosmological apparent horizon is timelike and inside
the cosmological event horizon, as shown in
Appendix~\ref{sec:friedmann}, 
the external Friedmann solution could only be matched to an internal self-similar solution at the cosmological event horizon. 
As shown in Appendix~\ref{sec:similarity_horizon_massless}, this horizon is a saddle, so there are just two 
isolated solutions which can be matched to 
the Friedmann background solution:
one is Friedmann itself and the other is a shock-wave solution. However, 
since we assume regularity of the cosmological event horizon, 
$z$, $y$ and $\beta$
must be the same in the shock-wave solution and
the exterior Friedmann background. The regularity therefore requires 
the same value of $\gamma$ and hence this similarity horizon
is identical to that of the exterior Friedmann solution. 
Since 
the interior can only be
the Friedmann solution itself,
we can conclude that there is no self-similar black 
hole solution interior to the Friedmann background.
This proof does not exclude the possibility of a self-similar black hole
in the asymptotically inflationary Friedmann background and this
possibility is now under investigation~\cite{mhc}.

\section{CONCLUSION}
In summary, we have shown that there is no self-similar 
black hole solution surrounded 
by an exact flat Friedmann solution for a massless scalar field
or a scalar field with a potential.  This extends the previous result obtained for a perfect
fluid with equation of state $p=k\epsilon$ ($0\le k \le 1$).
Indeed we can summarize this important result as a formal theorem.
\begin{The*}
Let $({\cal M}, g)$ be a spacetime which satisfies the 
following
four conditions:
(i) it is spherically symmetric;
(ii) it admits a homothetic Killing vector;
(iii) it contains a stiff fluid or a scalar field with non-negative 
potential;
(iv) it contains no singular hypersurface.
Let $({\cal M}, g)$ also satisfy at least one of the following 
two conditions:
(v) it coincides with the flat Friedmann solution
outside some finite radius;
(vi) it is asymptotic to the decelerating Friedmann
solution.
Then $({\cal M}, g)$ has no black-hole event horizon.
\end{The*}
This result disproves recent claims that PBHs could grow self-similarly
through accreting a quintessence field.
It also suggests that accretion onto black holes is suppressed by 
the cosmological expansion in a scalar field system.
This is consistent with numerical simulations, 
which show that PBHs with different initial sizes
eventually become much smaller than the 
particle horizon~\cite{hc2005b,hc2005c}.
Note that, for the massless case, it has already been shown analytically that the accretion rate is exactly
zero for a black hole whose size is exactly that of the cosmological apparent
horizon~\cite{hc2005b,hc2005c}. 

Although Bicknell and Henriksen's construction
shows that there are self-similar black hole solutions satisfying the necessary
energy conditions, such solutions are {\it ad hoc}
and probably not physically realistic. 
Certainly the PBH mass could grow 
self-similarly ($M\propto t$) only with very special matter fields. 
On the other hand, if we drop 
self-similarity condition,
there might still be black hole solutions which exhibit an interesting amount of accretion.
For example, kinematically self-similar 
solutions~\cite{coley1997,mh2005b} 
contain characteristic scales and this suggests consideration of
a wider class of matter fields.
In this case, one might have solutions in which the PBH grows as
$M\propto t^{1/\alpha}$ ($\alpha\ne 1$) for 
some similarity index $\alpha$.

\acknowledgments
TH and HM thank  H.~Kodama, K.~Nakamura, E.~Mitsuda and N.~Dadhich
for helpful discussions and comments. TH thanks T.~Nakamura
for helpful comments. This work was partly supported by the 
Grant-in-Aid for Young Scientists (B) 18740144 (TH) and 
18740162 (HM)
from the Ministry of Education, Culture, Sports,
Science and Technology (MEXT) of Japan. 
BJC thanks the Research Center for the Early Universe at 
Tokyo University for hospitality during this work.
The numerical and algebraic calculations were carried out
on machines at YITP in Kyoto University.

\appendix

\section{COMPARISON WITH OTHER WORK}
\label{sec:otherwork}

It is interesting to explain why the results of this paper disagree with those presented in various earlier papers.  First, we clarify the flaw in the claim by Lin et al.~\cite{lcf1976} that a self-similar black hole solution can be attached to the Friedmann particle horizon if the equation of state is stiff. This flaw was first pointed out by Bicknell and Henriksen~\cite{bh1978a} but our analysis has illuminated its nature. Lin et al. use different variables from those in Section~\ref{sec:stiff_fluid} but it is easy to express the results in our notation. They prove analytically that, as the similarity variable $X$ decreases, the function $V$ must reach a minimum below 1 and then rise again to 1 (cf. Fig.~\ref{fig:expected_V2}). They correctly point out that the pressure and velocity gradient diverge at this point but they fail to appreciate that the density function $E$ goes to zero and that the metric function $\nu$ goes to infinity, so that the surface $V^2=1$ is timelike rather than null. They therefore misinterpret this point as a black hole event horizon. They argue that the divergence of the pressure gradient is non-physical, since it can be removed by introducing an Eddington-Finkelstein-type coordinate, but do not notice that the pressure gradient also diverges at the cosmological particle horizon.  However, it is worth stressing that Lin et al. do correctly deduce the form of the solution up to the inner point where $V^2=1$.

Second, we point out the source of our disagreement with the claim of
Bean and Magueijo~\cite{bm2002} that black holes can accrete
quintessence fast enough to grow self-similarly. They consider a scalar
field in a potential of the form given by Eq.~(\ref{eq:potential}). 
By generalizing the analysis of Jacobson~\cite{jac1999}, which attaches a Schwarzschild solution to a cosmological background in which the field has the asymptotic value $\phi_c$, they claim that the energy flux through the event horizon is $T_{vv}=\dot{\phi_c}^2$, leading to a black hole accretion rate 
\begin{equation}
\frac{dM}{dt} = 16\pi M^2\dot{\phi_c}^2.
\label{eq:accretion}
\end{equation}
They correctly infer that only the kinetic energy of the scalar field is accreted, the scalar potential merely influencing the evolution of the asymptotic background field:
\begin{equation}
\phi_c =\frac{2}{\lambda \sqrt{8\pi}} \log\left(\frac{t}{t_0}\right)
\end{equation}
where $t_{0}$ is a constant of integration. The accretion rate therefore becomes
\begin{equation}
\frac{dM}{dt} = \frac{KM^2}{t^2}
\end{equation}
with $K=8/\lambda^2$, which can be integrated to give
\begin{equation}
M=\frac{M_0}{1-\displaystyle{\frac{KM_0}{t_0}\left(1-\frac{t_0}{t}\right)}}\;\; .
\label{eq:pbhmass}
\end{equation}
For $M_0 \ll K^{-1}t_0$, corresponding to an initial black hole mass
much less than that of the particle horizon, this implies negligible
accretion. However, for $M_0 = K^{-1}t_0$, Eq.~(\ref{eq:pbhmass}) predicts a self-similar solution, $M\propto t$, in which the black hole always has the same size relative to the particle horizon. This is the result which Bean and Maguiejo exploit. For $M_0 > K^{-1}t_0$, the mass diverges at a time
$t=t_{0}/[1-t_0/(KM_0)]$.

However, this analysis is exactly equivalent to that of Zeldovich and 
Novikov~\cite{zn1967}, which neglects the background cosmological expansion and which Carr and Hawking disproved. Bean and Magueijo caution that the Carr-Hawking proof only applies for a perfect fluid and may not be relevant if the scalar field has a potential. However, the result of this paper confirms that there is no black hole self-similar solution even in this case. (Bettwieser and Glatzel~\cite{bg1981} also question the Carr-Hawking analysis but they only use a general causality argument and do not give a detailed calculation.)
It is interesting to note that a more refined Newtonian analysis (but still neglecting the background expansion) implies that the critical black hole size for self-similar growth exceeds the scale associated with a separate closed universe condition for an equation of state parameter exceeding 0.6~\cite{hc2005a}. This also indicates that the Newtonian prediction cannot be correct. 

Finally, we comment on the paper by Custodio and Horvath~\cite{ch2005}. They also consider black hole accretion of a quintessence field with a scalar potential but, instead of Eq.~(\ref{eq:accretion}), obtain
 \begin{equation}
\frac{dM}{dt} = 27\pi M^2\dot{\phi_c}^2,
\end{equation}
where the numerical factor differs from that assumed in Eq.~(\ref{eq:accretion}) because of relativistic beaming. They disagree with Bean and Magueijo's choice of the function $\dot{\phi_c}$ on the grounds that it neglects the local decrease in the background scalar field resulting from the accretion.  For reasons which are not altogether clear, they therefore prefer to focus on a model in which the quintessence flux into the black hole is constant. This requires that the potential have a specific shape, which is not in fact of the form (\ref{eq:potential}) required by self-similarity. This leads to the mass evolution
 \begin{equation}
M=\frac{M_0}{1-\displaystyle{\frac{\bar{K}M_0}{t_0}}(t-t_0)} \, ,
\end{equation}
where $\bar{K}$ is a constant related to the (fixed) flux. Again this is
similar to the Zeldovich-Novikov formula. The mass diverges 
at a time $t=t_{0}[1+1/(\bar{K}M_0)]$, just as Eq.~(\ref{eq:pbhmass})
does for $M_0 > K^{-1}t_0$, and they attribute this unphysical feature 
to the fact that the assumption of constant mass flux fails. 
They therefore consider alternative models in 
which the flux decreases as a power of time. 

Although we agree with their conclusion that the Bean-Magueijo argument is
wrong, we would claim that Custodio and Horvath have not identified the
more significant reason for its failure, namely that it neglects the
background cosmological expansion. In general, they find that the
increase in the black hole mass is small unless its initial value is
finely tuned. However, according to the result in this paper, even the
latter claim is misleading.

\section{Global structure of Friedmann solution}
\label{sec:friedmann}
Since we are considering the possibility of black holes 
in a Friedmann background,
we briefly review the global structure of 
the flat Friedmann solution. This
has the metric
\begin{equation}
ds^{2}=-dt^{2}+a^{2}(t)[d\chi^{2}+\chi^{2}(d\theta^{2}+\sin^{2}d\phi^{2})]
\end{equation}
with the scale factor $a(t)=C t^{q}$,
where $C$ and $q$ are positive constants.
The domain of the variables is $0<t<\infty$ and $0\le\chi<\infty$.
The big bang singularity is $t=0$ and the regular centre is $\chi=0$.
We have $q=2/[3(1+k)]$ for a perfect fluid with 
$p=k\epsilon$ ($k> -1$), $q=1/3$ for a 
massless scalar field ($k=1$) and $q=2/\lambda^{2}$
for a scalar field $\phi$
with the exponential potential $V=V_{0}e^{-\sqrt{8\pi}\lambda\phi}$.
The cosmic expansion is decelerated for $0<q<1$, accelerated for $q>1$
and has constant speed for $q=1$.

In terms of the conformal time $\eta$, defined by $a d\eta=dt$,
we have
$a\propto |\eta |^{q/(1-q)}$ for $q \neq 1$ and
$a\propto e^{C\eta}$ for $q=1$.
The range of $\eta$ is 
$0<\eta<\infty$ for $0<q<1$, $-\infty<\eta<\infty$
for $q=1$ and $-\infty < \eta <0$ for $q>1$.
Defining new variables $\hat{t}$ and $\hat{r}$ by
\begin{eqnarray}
\eta &=&\frac{1}{2}\left[\tan
		    \left(\frac{\hat{t}+\hat{r}}{2}\right)+
\tan\left(\frac{\hat{t}-\hat{r}}{2}\right)\right], \\  
\chi&=&\frac{1}{2}\left[\tan
		    \left(\frac{\hat{t}+\hat{r}}{2}\right)-
\tan\left(\frac{\hat{t}-\hat{r}}{2}\right)\right],
\end{eqnarray}
we can draw the Penrose diagram for each case. 
Through this transformation,
the big bang singularity $t=0$ and the regular centre $\chi=0$
are transformed respectively to
$\hat{t}=0$ and $\hat{r}=0$ for $0<q<1$
and to 
$\hat{t}-\hat{r}=-\pi$ and $\hat{r}=0$ for $q\ge 1$.
It can be also shown~\cite{senovilla1997} that future null infinity is 
given by $\hat{t}+\hat{r}=\pi$ for $0<q\le 1$
and $\hat{t}=0$ for $q>1$.
The big bang singularity is spacelike for $0<q<1$
but null for $q \ge 1$. 

\begin{figure*}[htbp]
\begin{tabular}{cc}
\subfigure[$0<q<1/2$]
{\label{fig:hard_flat_friedmann} 
\includegraphics[width=6cm]{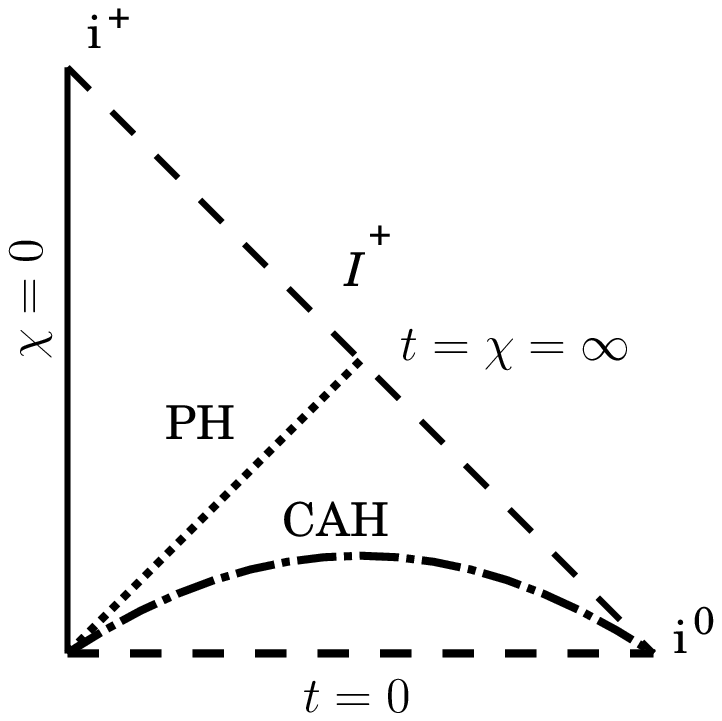}}&
\subfigure[$q=1/2$]{\label{fig:rad_flat_friedmann} \includegraphics[width=6cm]{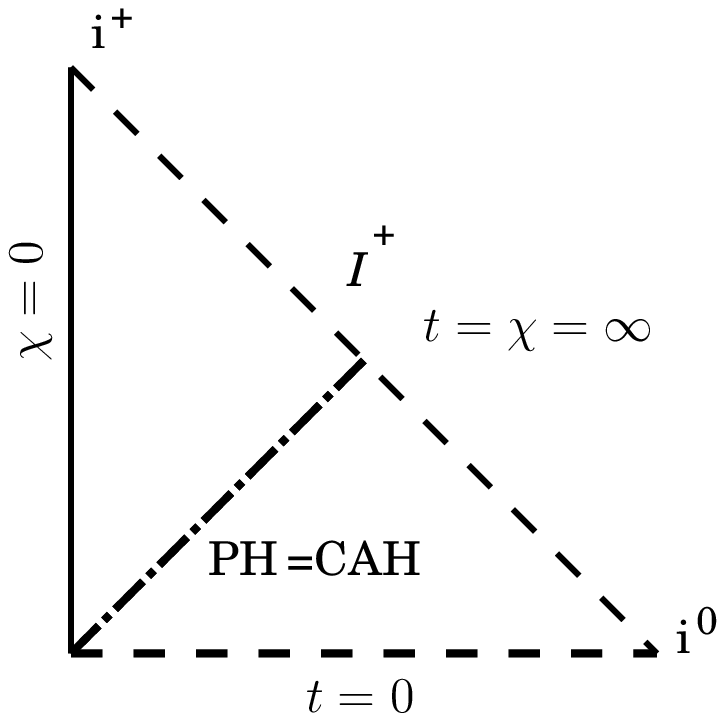}}\\
\subfigure[$1/2<q<1$]
{\label{fig:soft_flat_friedmann} \includegraphics[width=6cm]{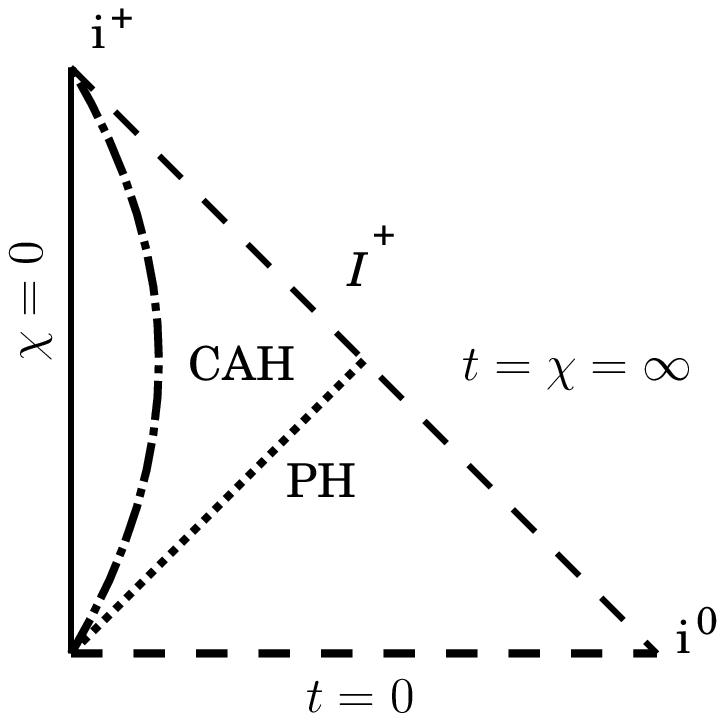}}&
\subfigure[$q=1$]{\label{fig:equivelocity_flat_friedmann}\includegraphics[height=7cm]
{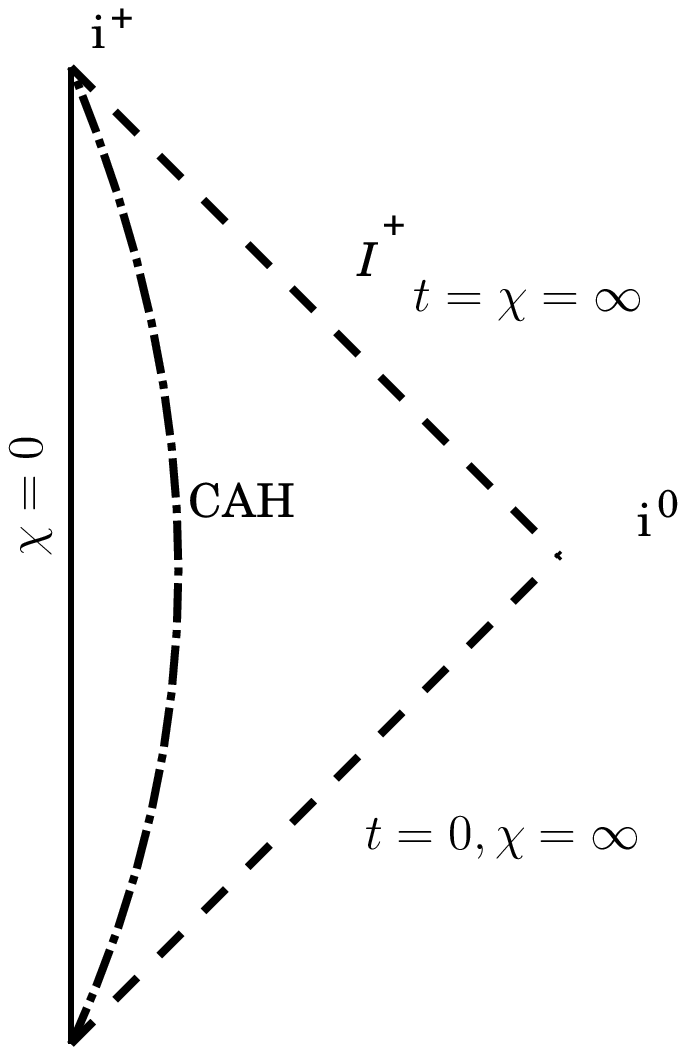}}\\
\subfigure[$q>1$]
{\label{fig:accelerated_flat_friedmann} 
\includegraphics[width=6cm]{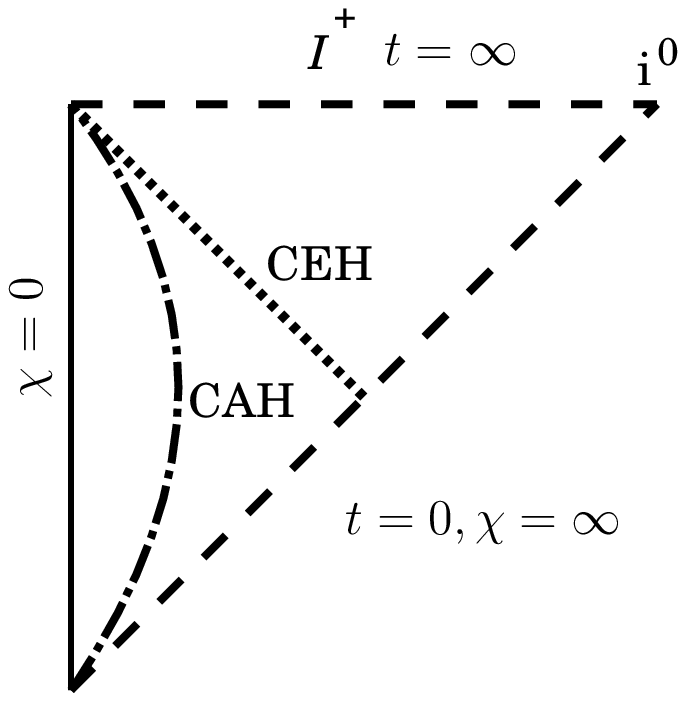}}& \\
\end{tabular}
\caption{\label{fig:penrose_flat_friedmann}
The causal structure of the Friedmann universe with the scale factor
$a \propto t^{q}$ for different values of $q$.
The particle horizon (PH), cosmological apparent horizon (CAH)
and cosmological event horizon (CEH) are indicated.}
\end{figure*}

When we introduce the double null coordinates 
\begin{equation}
u=\eta-\chi,\quad 
v=\eta+\chi, 
\end{equation}
the line element takes the form
\begin{equation}
ds^{2}=a^{2}[-dudv+\chi^{2}(d\theta^{2}+\sin^{2}\theta d\phi^{2})],
\end{equation}
where 
\begin{equation}
a^{2}=C^{2/(1-q)}\left[(1-q)\left(\frac{u+v}{2}\right)\right]^{2q/(1-q)}
\end{equation}
for $q \neq 1$ and
\begin{equation}
a^{2}=C^{2} e^{C (u+v)},
\end{equation}
for $q=1$. 
The area radius $r$ is given by $r=a(v-u)/2$.  
The Misner-Sharp mass $m$ is written in this metric as 
\begin{equation}
m=\frac{r}{2}\left(1+\frac{4r_{,u}r_{,v}}{a^{2}}\right).
\end{equation}
Then
\begin{equation}
\frac{2m}{r}=\left(\frac{q}{1-q}\right)^{2}\left(\frac{v-u}{v+u}\right)^{2},
\end{equation}
for $q \neq 1$ and 
\begin{equation}
\frac{2m}{r}=\frac{C^{2}}{4}(v-u)^{2}
\end{equation}
for $q=1$. The past trapping horizon, which coincides with 
the cosmological apparent horizon, is given by $2m/r=1$, or 
\begin{equation}
u=(2q-1)v
\end{equation}
for $q \neq 1$ and
\begin{equation}
v-u=\frac{2}{C},
\end{equation}
for $q=1$. Therefore, 
the cosmological apparent horizon is 
spacelike for $0<q<1/2$, null for $q=1/2$
and timelike for $q>1/2$.
Figure~\ref{fig:penrose_flat_friedmann} shows five different Penrose 
diagrams, depending on the value of $q$.
There is a particle horizon, given by $u=0$, only for $0<q<1$.
There is a cosmological event horizon, given by $v=0$, only for $q>1$.
The flat Friedmann solution for $q=1$ has neither a
particle horizon nor a cosmological event horizon.

\section{Friedmann solution as a self-similar spacetime}
\label{sec:friedmann_self-similar}
Although these flat Friedmann solutions are self-similar in the sense
of homothety, it is not obvious in the present form.
Using double-null coordinates, the standard form of 
spherically symmetric self-similar spacetimes is given by
\begin{equation}
ds^{2}=-{\tilde a}^{2}(\sigma)d\tilde{u}d\tilde{v}
+\tilde{u}\tilde{v}\tilde{b}^{2}(\sigma)(d\theta^{2}
+\sin^{2}\theta d\phi^{2}),
\end{equation}
where $\sigma=\tilde{u}/\tilde{v}$ is the similarity variable.
Since the metric with the double null 
coordinates $u$ and $v$ does not satisfy this form,
we transform to new double null coordinates $\tilde{u}$
and $\tilde{v}$.

For $q \neq 1$,
$\tilde{u}$ and $\tilde{v}$ are given by
\begin{equation}
 u=\pm |\tilde{u}|^{1-q},\quad v=\pm
 |\tilde{v}|^{1-q},
\end{equation}
where the signs are chosen so that the sign of 
$\tilde{u}$ ($\tilde{v}$) coincides 
with the sign of $u$ ($v$) and we can show $-1<\sigma\le 1$.
Then the metric functions become
\begin{eqnarray}
\tilde{a}^{2}&=&\mbox{const}\times 
\left||\sigma|^{-(1-q)/2}\pm |\sigma|^{(1-q)/2}\right|^{2q/(1-q)}, \\
\tilde{b}^{2}&=&\mbox{const}\times 
\left||\sigma|^{-(1-q)/2}\pm |\sigma|^{(1-q)/2}\right|^{2q/(1-q)}
\nonumber \\
&& \times ||\sigma|^{-(1-q)}\mp |\sigma|^{(1-q)}|^{2},
\end{eqnarray}
where the upper and lower signs correspond
to the positive and negative values of $\sigma$, respectively.
For $q=1$,
\begin{equation}
\tilde{u}=e^{Cu}, \quad
\tilde{v}=e^{Cv}
\end{equation}
and the metric functions become
\begin{eqnarray}
\tilde{a}^{2}&=&\mbox{const}, \\
\tilde{b}^{2}&=&\mbox{const}\times (\ln \sigma)^{2},
\end{eqnarray}
where $0<\sigma\le 1$.

The similarity surface is generated by a homothetic
Killing vector.
When this is null, it is called a similarity horizon 
and this does not depend on the choice of similarity
variable.
The induced metric on the similarity surface 
$\sigma=\mbox{const}$ is 
\begin{eqnarray}
ds^{2}&=&-\tilde{a}^{2}\sigma^{-1} d\tilde{u}^{2} + 
\tilde{u}\tilde{v}\tilde{b}^{2}(d\theta^{2}
+\sin^{2}\theta d\phi^{2})\nonumber \\
&=& - \tilde{a}^{2}\sigma d\tilde{v}^{2}+
\tilde{u}\tilde{v}\tilde{b}^{2}(d\theta^{2}
+\sin^{2}\theta d\phi^{2}).
\end{eqnarray}
We provide two expressions because one of them may
become invalid when one of the coordinates becomes degenerate. 
For $q \neq 1$, the only
similarity horizon is $\sigma=0$. This is outgoing and null for $0<q<1$,
corresponding to the particle horizon $u=0$, and ingoing and null for $q>1$, corresponding to
the cosmological event horizon $v=0$.
For $q=1$, there is no similarity horizon and the
similarity surfaces are $v-u=\mbox{const}$, which are timelike.
The similarity horizon is important
because it is the matching surface between the external Friedmann 
background and the internal self-similar solution.

\section{Analysis of similarity horizon for the massless case}
\label{sec:similarity_horizon_massless}

We introduce a new independent variable $u$ defined by
\begin{equation}
\frac{d \xi}{d u}=1-2z,
\label{defineu}
\end{equation} 
where the similarity horizon $\xi=\xi_{\rm s}$ corresponds to $u = \pm \infty $ and
is an equilibrium point. From Eqs.~(\ref{eq:zdot}) -- (\ref{eq:gammadot}), we can write the ODEs near $\xi_{\rm s}$ in the form:
\begin{eqnarray}
\frac{dz}{du}&=&z(1-2z)[2-y^{-1}], \\
\frac{dy}{du}&=&(1-2z)[1-(1+4\pi\gamma^{2})y], \\
\frac{d\gamma}{du}&=&2\kappa z-\gamma (y^{-1}-2z), 
\end{eqnarray}
together with the constraint given by Eq.~(\ref{eq:const2}).
We now write the asymptotic solution near 
$\xi_{\rm s}$ as
\begin{equation}
z=\frac12(1+x_{1}) , 
~y=y_{\rm s}(1+x_{2}), 
~\gamma=\gamma_{\rm s}(1+x_{3}),
~\xi=\xi_{\rm s}+x_{4}, 
\end{equation}
where $y_{\rm s}$ and $\gamma_{\rm s}$ are given by Eq.~(\ref{ygamma}),
and regard $(x_{1},x_{2},x_{3},x_{4})$ as defining a 4-vector ${\bf x}$. 
The linearised ODEs can be written in the matrix form
\begin{equation}
\frac{d}{du}{\bf x}={\bf A}{\bf x}
\end{equation}
where 
\begin{equation}
{\bf A}=
\left(\begin{array}{cccc}
\alpha-1 & 0 & 0 & 0\\
\alpha^{-1}-\alpha  & 0 & 0 & 0 \\
1+\alpha & 1+\alpha & -\alpha & 0 \\
-1 & 0 & 0 & 0
\end{array}\right)
\end{equation}
and $\alpha\equiv 4\pi\kappa^{2}$.
This matrix generically has only two 
non-zero eigenvalues:
\begin{equation} 
\lambda_{1}=-1+\alpha,\quad \lambda_{2}=-\alpha.
\end{equation}
These are associated with two eigenvectors
\begin{equation}
{\bf e}_{1}=
\left(\begin{array}{c}
\alpha(1-2\alpha)(1-\alpha) \\
-(1-2\alpha)(1-\alpha)(1+\alpha) \\
(1-\alpha)(1+\alpha) \\
\alpha(1-2\alpha)
\end{array}\right), \quad
{\bf e}_{2}=
\left(\begin{array}{c}
0 \\
0 \\
1 \\ 
0 
\end{array}\right),
\end{equation}
and the two associated values of $\dot{\gamma}_{\rm s}=(\gamma x_3/x_4)_{\rm s}$ are given by Eq.~(\ref{gammadot}).
In particular, there is a solution belonging to 
the eigenvector ${\bf e}_{2}$ which has the form
\begin{equation}
z=\frac{1}{2}, \quad y=y_{\rm s}, \quad \gamma=\gamma_{\rm s}(1\pm e^{-\alpha u}), 
\quad \xi=\xi_{\rm s},
\label{eq:shock_wave}
\end{equation}
and which we call a shock-wave solution.
Note that the equilibrium point is a node for $\lambda_1 <0$ and $\lambda_2 <0$, with the primary direction being associated with the higher value, and a saddle for $\lambda_1 >0$ and  $\lambda_2 <0$. 

\section{Similarity horizon for the massless case revisited}
\label{sec:follow_brady}
Here we revisit Brady's non-linear analysis of the similarity horizon
for a massless scalar field~\cite{brady1995}.
We introduce new variables $\zeta$ and $\eta$ such that
\begin{eqnarray}
z=\frac12+\zeta , \quad y=y_{\rm s}+\eta,\label{eta} \quad y_{\rm s}=
\frac{1}{1+4\pi\kappa^2},
\end{eqnarray}
where $\zeta=\eta=0$ corresponds to the similarity horizon and
only $\eta/\zeta\le 0$ is allowed. 
When $\zeta$ and $\eta$ are sufficiently small,
Eqs.~(\ref{eq:zdot})-(\ref{eq:const2}) reduce to
\begin{eqnarray}
\gamma&=&-\kappa\pm \frac{1+4\pi\kappa^{2}}{\sqrt{8\pi}}
\sqrt{-\frac{\eta}{\zeta}}, \label{bb10} \\
\frac{d\eta}{d\zeta}&=&\pm \frac{2\sqrt{8\pi \kappa^{2}}}{1-4\pi\kappa^{2}}
\sqrt{-\frac{\eta}{\zeta}}+\frac{1+4\pi\kappa^{2}}{1-4\pi\kappa^{2}}
\frac{\eta}{\zeta} \, .
\label{bb14}
\end{eqnarray}
Here and throughout Appendix~\ref{sec:follow_brady} 
the upper and lower signs correspond
to the positive and negative branches in Eq.~(\ref{bb10}).
These equations can immediately be integrated to give
\begin{eqnarray}
\sqrt{\frac{-\eta}{\zeta}}&=&\pm \frac{1}{\sqrt{2\pi}\kappa}
+C |\zeta|^{4\pi\kappa^{2}/(1-4\pi\kappa^{2})}, 
\label{eq:asymptotic_solution1}\\
\gamma&=&\frac{1}{4\pi\kappa}\pm \frac{C(1+4\pi\kappa^{2})}{\sqrt{8\pi}}
|\zeta|^{4\pi\kappa^{2}/(1-4\pi\kappa^{2})},
\label{eq:asymptotic_solution2}
\end{eqnarray}
where $C$ is an integration constant.
If $\zeta<0$ and $\eta>0$,
Eq.~(\ref{eq:asymptotic_solution1}) with the upper sign 
coincides with Brady's solution (Eq.~(C6) in~\cite{brady1995}),
except for the exponent of the 
second term on the right-hand side, which may be a typographical error.
The negative-branch solutions do not cross the similarity 
horizon $\zeta=\eta=0$.

The solution with the upper sign has
the following behaviour.
If $C=0$, it always crosses the similarity horizon.
For $0<4\pi\kappa^{2}<1$, there is a one-parameter 
family of solutions which cross the similarity horizon,
because the second term on the right-hand side of
Eqs.~(\ref{eq:asymptotic_solution1}) and
(\ref{eq:asymptotic_solution2}) goes to zero as $\zeta\to 0$.
For $4\pi\kappa^{2}>1$, however, the solution with $C=0$
is isolated because $\eta $ will not go to zero 
as $\zeta \to 0$ if $C\ne 0$.
If we consider the derivatives along solutions with $C\ne 0$,
because of the second term,
$d\gamma/d\zeta $ is diverging for $ 0<4\pi\kappa^{2}<1/2$
and finite for $1/2 <4\pi\kappa^{2}<1$. This reflects 
the behaviour of sound-wave solutions with $C\ne 0$, which 
is discussed in Section~\ref{subsec:structure_similarity_horizon}
using the linearised ODEs.
Note that the shock-wave solution, which is given 
by Eq.~(\ref{eq:shock_wave}), now has 
$\eta=\zeta=0$ but changing $\gamma$.

\section{Analysis of similarity horizon for the potential case}
\label{sec:similarity_horizon_potential}
Using the variable $u$ defined by Eq.~(\ref{defineu}),
the system of ODEs near the similarity horizon has the form:
\begin{eqnarray}
\frac{dz}{du}&=&z(1-2z)[2-y^{-1}(1-\beta)], \\
\frac{dy}{du}&=&(1-2z)[1-\beta-(1+4\pi\gamma^{2})y], \\
\frac{d\gamma}{du}&=&2\kappa z-\gamma[y^{-1}(1-\beta)-2z]+\frac{\beta}
{4\pi\kappa y}, 
\label{gammareg}\\
\frac{d\beta}{du}&=&2(1-2z)\beta(1+\gamma/\kappa), 
\end{eqnarray}
together with the constraint
\begin{equation}
1-\beta=y[(1+4\pi\kappa^{2})-4\pi(\gamma+\kappa)^{2}(1-2z)].
\label{constraint}
\end{equation}
The nature of the equilibrium point
can again be classified by the behaviour of solutions there.
We put 
\begin{eqnarray*}
z&=&\frac12(1+x_1),
y=y_{\rm s}(1+x_{2}),
\gamma=\gamma_{\rm s}(1+ x_{3}),\\
\xi&=&\xi_{\rm s}+x_{4},
\beta=\beta_{\rm s}(1+ x_{5}), 
\end{eqnarray*}
where $x_{1}$ to $x_{5}$ are
regarded as components of a 5-vector 
${\bf x}$.
From the regularity condition at the similarity horizon associated with Eq.~(\ref{gammareg}) and the 
constraint equation (\ref{constraint}), we obtain 
Eq.~(\ref{eq:regular_similarity_horizon_potential}),
so that the similarity horizon is parametrized by the values of 
$\kappa$ and $\beta_{\rm s}$.

We now write the linearised 
ODEs in the form:
\begin{equation}
\frac{d}{du}{\bf x}={\bf A}{\bf x},
\end{equation}
where the matrix ${\bf A}$ is given by
\begin{equation}
{\bf A}=
\left(\begin{array}{ccccc}
\alpha-1 & 0 & 0 & 0 & 0 \\
A_{21} & 0 & 0 & 0 & 0 \\
A_{31} & A_{32} & -\alpha & 0 & A_{35} \\
-1 & 0 & 0 & 0 & 0 \\
A_{51} & 0 & 0 & 0 & 0 
\end{array}\right)
\end{equation}
with
\begin{eqnarray}
A_{21}&=&\frac{(1+\alpha)[\beta_{\rm s}(1+\alpha^2)+\alpha(1-\alpha)]}
{\alpha^3(1-\beta_{\rm s})^2} \nonumber \\
&& \times [(1-\alpha) \beta_{\rm s}+\alpha],\\
A_{31}&=&A_{32}=\frac{(1+\alpha)[(1-\alpha)\beta_{\rm s}+\alpha]}{\beta_{\rm s}+\alpha},\\
A_{35}&=&\frac{\beta_{\rm s}(1+\alpha)[(1-\alpha)\beta_{\rm s}+2\alpha]}{(1-\beta_{\rm s})
(\beta_{\rm s}+\alpha)},\\
A_{51}&=&-\frac{2(1+\alpha)[(1-\alpha)\beta_{\rm s}+\alpha]}{\alpha^2(1-\beta_{\rm s})},
\end{eqnarray}
where $\alpha \equiv 4\pi\kappa^{2}$. 
This matrix has three zero eigenvalues and two 
generically non-zero ones, $\lambda_{1}=-1+\alpha$ and $\lambda_{2}
=-\alpha$.
Two of the eigenvectors associated with zero eigenvalues do not 
satisfy the constraint equation and so are ruled out.
The eigenvectors associated with the eigenvalues $\lambda_{1}$,
$\lambda_{2}$ are respectively
\begin{equation}
{\bf e}_{1}=
\left(\begin{array}{c}
1-\alpha \\
(e_1)^{2} \\
(e_1)^{3} \\ 
1 \\ 
(e_1)^{5} 
\end{array}\right),~~~
{\bf e}_{2}=
\left(\begin{array}{c}
0 \\
0 \\
1 \\ 
0 \\
0 
\end{array}\right),
\end{equation}
where
\begin{eqnarray}
(e_1)^{2}&=&-\frac{(1+\alpha)[(1+\alpha^2)\beta_{\rm s}+\alpha(1-\alpha)]}
{\alpha^3(1-\beta_{\rm s})^2} \nonumber \\
&& \times [(1-\alpha)\beta_{\rm s}+\alpha],\\
(e_1)^{3}&=&\frac{(1+\alpha)[(1-\alpha)\beta_{\rm s}+\alpha]Q}{\alpha^3(1-\beta_{\rm s})^2
(1-2\alpha)(\beta_{\rm s}+\alpha)},\\
Q&\equiv&-(1-\alpha)(\alpha^2+\alpha-1)\beta_{\rm s}^2 \nonumber \\ 
&& -2\alpha(\alpha^2+2\alpha-1)\beta_{\rm s} +\alpha^2(1-\alpha), \\
(e_1)^{5}&=&\frac{2(1+\alpha)[(1-\alpha)\beta_{\rm s}+\alpha]}{\alpha^2(1-\beta_{\rm s})}.
\end{eqnarray}
These eigenvectors lie on the constraint surface.

\end{document}